\pdfoutput=1
\pdfoutput=1
\pdfoutput=1
\pdfoutput=1
\pdfoutput=1
\documentclass[12pt, draftclsnofoot, onecolumn]{IEEEtran}
\normalsize

\usepackage{enumerate}
\usepackage{color}

\usepackage{cite}

\usepackage{float}
\newfloat{figtab}{htb}{fgtb}
\makeatletter
\newcommand\figcaption{\def\@captype{figure}\caption}
\newcommand\tabcaption{\def\@captype{table}\caption}
\makeatother

\usepackage{graphicx}
\usepackage{booktabs}
\usepackage{algorithm}
\ifCLASSINFOpdf
\else
\fi

\usepackage{amsthm}
\usepackage{amssymb,amsfonts}
\usepackage{amsfonts,balance}

\usepackage[cmex10]{amsmath}

\newtheorem{theorem}{Theorem}

\usepackage{algorithmic}

\usepackage{array}
\usepackage{multirow,makecell}
\usepackage[caption=false,font=footnotesize]{subfig}

\usepackage{verbatim}

\begin{document}
%
\title{Learning-based Prediction, Rendering and Association Optimization for MEC-enabled Wireless Virtual Reality (VR) Network}
%
%
%

\author{Xiaonan~Liu,~\IEEEmembership{Student Member,~IEEE,}
        Yansha~Deng,~\IEEEmembership{Member,~IEEE,}\\}

\maketitle

\begin{abstract}
Wireless-connected Virtual Reality (VR) provides immersive experience for VR users from anywhere at anytime. However, providing wireless VR users with seamless connectivity and real-time VR video with high quality is challenging due to its requirements in high Quality of Experience (QoE) and low VR interaction latency under limited computation capability of VR device. To address these issues, we propose a MEC-enabled wireless VR network, where the field of view (FoV) of each VR user can be real-time predicted using Recurrent Neural Network (RNN), and the rendering of VR content is moved from VR device to MEC server with rendering model migration capability. Taking into account the geographical and FoV request correlation, we propose  centralized and distributed decoupled Deep Reinforcement Learning (DRL) strategies to maximize the long-term QoE of VR users under the VR interaction latency constraint. Simulation results show that our proposed MEC rendering schemes and DRL algorithms substantially improve the long-term QoE of VR users and reduce the VR interaction latency compared to rendering at VR devices.
\end{abstract}




%
\IEEEpeerreviewmaketitle

\section{Introduction}
With the development of virtual reality (VR) technology, the interactions between VR users and their world will be revolutionized. VR can connect users across global communities within highly immersive virtual worlds that breaks geographical boundaries. This vision has inspired the commercial release of various hardware devices, such as Facebook Oculus Rift \cite{VR_device}. Indeed, it is anticipated that 99 million VR devices are needed in 2021 \cite{VR2021}, and that the market will reach 108 billion dollars by then \cite{VRgrowth}. However, the poor user experience provided by traditional computer-supported VR devices constrains the type of activities and experience of the VR user. One of the main barriers of the wired connected VR devices is the limited mobility of VR users. To overcome this disadvantage, wireless connected VR devices can be potential solution in providing ubiquitous user experiences from anywhere at anytime, and also can unleash plenty of novel VR applications \cite{Hu1}. Nevertheless, there are some unique challenges in wireless VR system that does not exist in wired VR system and traditional wireless video transmission system as identified in \cite{Hu1}. This includes how to provide seamless and real-time VR video with high quality through unstable wireless channels, solve handover issues when VR users are in mobility, and support the asymmetric and coupled trafﬁc in the uplink and downlink transmission \cite{Hu1, VR_challenge1,VR_challenge2,VR_system}. 

There are growing research interests in wireless VR networks. In \cite{VR1}, the echo state network (ESN) was proposed to solve the resource block allocation problem in both uplink and downlink wireless VR transmission to maximize the average quality of service (QoS) of VR users. Extending from \cite{VR1}, the authors in \cite{VR2} optimized the resource block allocation using ESN to maximize  the success transmission probability accounting for the VR user data correlation. In \cite{VR6}, through caching part of VR video frames in advance at the server and computing certain post-processing procedures on demand at the mobile VR device, the joint caching and computing optimization problem was formulated to minimize the average required transmission rate to reduce communication bandwidth.

The aforementioned VR research  mainly considered the VR video rendering occurs at the VR device side. For real-time interactive VR applications,  the latest 2D  video content needs to be first delivered to the VR device via wired/wireless communication, and then rendered to 3D VR video locally. In reality, human wearing VR device only watches a portion of observable visual world at any given time, which is so called Field of View (FoV).  Rendering the full 360 degree video in real-time can be costly both for downlink transmission and computation.  One potential solution is to only render the requested FoV each time based on the uplink tracking information of VR users' motion, including head and eye movements.  According to \cite{building_360}, the data size of the rendered FoV is 75$\%$ of that of the stitched 2D image, which means that the size of data to be delivered via downlink transmission can be reduced  by 25$\%$ compared to delivering the stitched 2D images.

Rendering real-time VR videos with high quality demands the computing unit with high processing ability, so that the rendering latency can be reduced, unfortunately, the computation ability and battery capacity of wireless VR devices are limited. Recently, mobile edge computing (MEC) has emerged to push mobile computing and network control to network edge, so as to enable computation-intensive and latency-critical applications at the resource-limited mobile devices, which promise dramatic reduction in latency and energy consumption \cite{Jun_zhang_mec}. 

Shifting the FoV rendering task from VR device to MEC server can not only alleviate the computation requirement at the VR device, but also potentially decrease the VR interaction latency especially for those VR users in the same virtual VR environment requesting the same FoV. With MEC multicast to a group of VR users selecting the same FoV, and unicast to single VR user selecting unique FoV, the downlink transmission cost of network can be further decreased. In practice, each MEC has different computation capability, it would be interesting to explore if we can obtain further gain for  multiple VR user groups selecting the same FoV but different MECs by performing rendering at only one MEC, and then migrate the rendered FoV wired to other MECs, namely, rendered FoV migration.

As pointed out by \cite{Hu1}, the Quality of Experience (QoE) of VR transmission is substantially different from that of conventional video transmission, due to its unique requirements in the VR interaction latency, and asymmetric uplink and downlink data rates. Motivated by above, in this paper, we focus on optimizing the QoE of VR users with interactive VR applications in a MEC-enabled wireless VR networks, and we develop a decoupled learning strategy to efficiently optimize the QoE in wireless VR system, which can improve the training efficiency \cite{Jiang1,Jiang2}. The main contributions can be summarized as follows:

\begin{itemize}
    \item We propose a MEC-enabled wireless VR networks, where the field of view (FoV) of each VR user can be real-time predicted, and the rendering of VR content is moved from VR device to MEC server with rendering model migration capability. 
    \item With the aim of optimizing the long-term QoE of VR users, we propose a decoupled learning strategy. This strategy decouples the optimization by separately resolving two-sub tasks, which are FoV prediction and rendering MEC association with the help of Recurrent Neural Network (RNN) predictor and Deep Reinforcement Learning (DRL) algorithms, respectively. 
    \item In order to capture the complex dynamics of the FoV request from each VR device, we propose the RNN model based on Gated Recurrent Unit (GRU) architecture at the central controller to predict the requested  FoV in the current time slot based on those in previous time slots sent via uplink. Our results shown that our proposed FoV prediction based on GRU
    achieves $96\%$ in prediction accuracy.  
    \item Accounting for the geographical and  predicted FoV request correlation, we propose  centralized and distributed decoupled Deep Reinforcement Learning (DRL) strategies based on Deep Q-Network (DQN) and Actor Critic (AC) \cite{centralize1,centralize2,distribute1,distribute2,distribute3,distribute4} to maximize the long-term QoE of VR users, via determining optimal association between MEC and VR user group, and  optimal rendering MEC for model migration. By comparing with non-learning based nearest MEC association, our results on centralized and distributed DQN shown substantial gain in both QoE and VR interaction latency. Interestingly, the rendering model migration further improve these gains.
\end{itemize}

The rest of this paper is organized as follows. The system model and problem formulation are proposed in Section II. RNN-based FoV prediction and DRL-based MEC rendering, migration and association scheme are presented in Section III. The simulation results and conclusions are described in Section IV and Section V, respectively.

\begin{figure}[ht]
    \centering
    \includegraphics[width=3.0in]{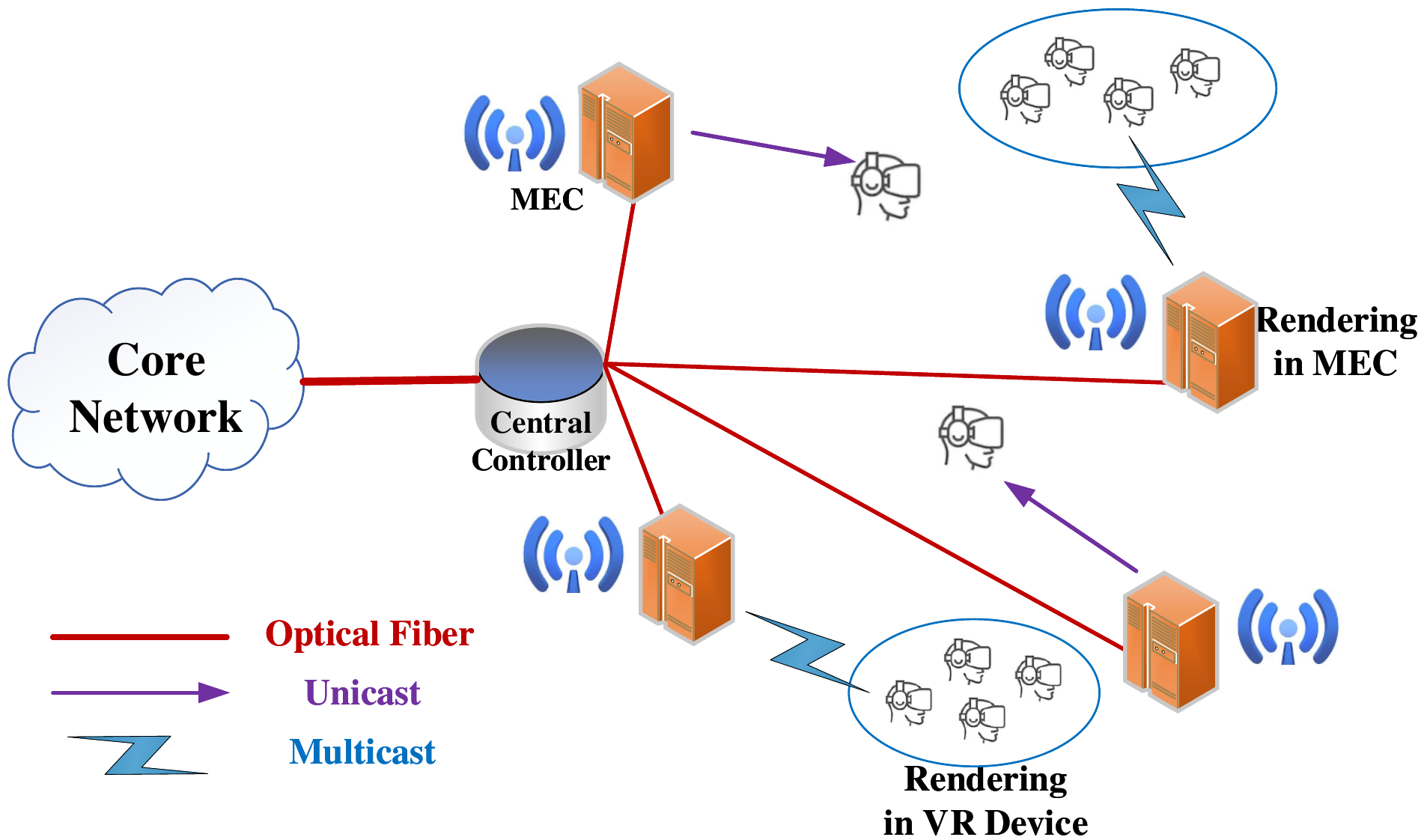}
    \caption{Wireless VR system in cellular network.}
    \label{basic_modules}
\end{figure}

\section{System Model and Problem Formulation}
We consider a wireless VR system where multiple MECs are connected to a central controller through a fiber link and serves $K_{\text{VR}}$ VR users via wireless links, as shown in Fig. 1. The central controller is connected to the core network via fiber, and can fetch real-time 2D videos without distortion from the core network.

\subsection{System Model}
Our MEC-enabled wireless VR system is consist of  four main parts, including FoV selection and prediction, uplink transmission, FoV rendering, and multi-group multicast and unicast downlink transmission. 
\subsubsection{FoV Selection and Prediction} When VR users enjoy the VR video, because of the restricted area of vision in human eye, they usually watch only a portion of the VR video, namely, the Field of View (FoV) \cite{Bao1,ERP}, which specifies a $150^{\circ}\times 135^{\circ}$ (i.e., diagonal $200^{\circ}$) FoV requirement. As the FoV is part of the actual rendered 3D VR video, the MEC benefits from obtaining the tracking information
related to the viewport of the VR user, and uses the video characteristics such as projection and mapping
formats to generate FoV. If the VR system could know the required FoV before transmission, it could render and deliver the FoV to VR users in advance, which can decrease the latency.

\begin{figure}[!t]
    \centering
    \includegraphics[width=3.0 in]{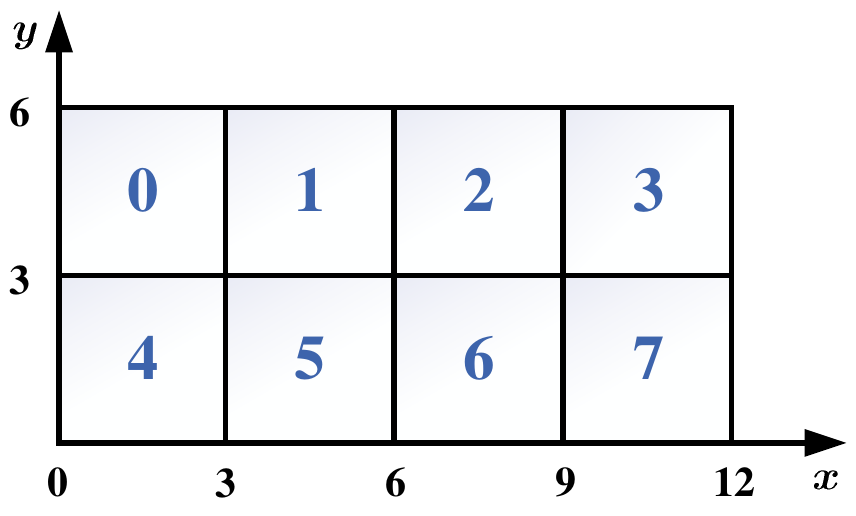}
    \caption{Brownian motion for FoV selection.}
    \label{basic_modules}
\end{figure}

Let us first denote the total number of FoVs of a VR virtual environment as $N_{\text{FoV}}$. When VR users enjoy the 360-degree video, they may randomly select the same or different FoVs in the continuous time slots. The Brownian motion can be used to model the eye movement of each VR user corresponding to different FoVs over time slots \cite{brownian3}. To capture the FoV selection in the 3D VR scenario close to reality, we map the 3D VR view into a large 2D view with $N_{\text{FoV}}$ FoVs in each time. When the VR user's eyes move inside the cubic in Fig. 2, the corresponding FoV will be selected. According to Brownian motion \cite{brownian1, brownian2}, the eye movement of the VR user at the $t$th time slot can be modeled by an independent Gaussian distribution with variance $2D(t)$ and zero mean $\mathcal{N}(0, 2D(t))$. Thus, the eye movement of the $k$th VR user in a mapped 2D VR view at the $t$th time slot can be expressed as
\begin{equation}
    \bigtriangleup S_{k}(t) = \{\mathcal{N}(0, 2D_k(t)), \mathcal{N}(0, 2D_k(t))\}.
\end{equation}
At one time slot, the selected FoV of the $k$th VR user is observed, which can be used for FoV prediction in the next time slot. For example, we assume that there are 8 2D FoVs in a VR virtual environment, as shown in Fig. 2. If the VR user selects the 2th FoV at a certain time slot, in the next time slot, it will select one of FoVs among the $1, 2, 3, 5, 6, 7$th FoVs. If the VR user selects a boundary FoV, such as the 0th FoV at a certain time slot, it can choose one FoV from the $0, 1, 4, 5$th FoVs in the next time slot. 

Based on the historical FoV selection in the previous time slots, the wireless VR system can predict the requested FoV of the $k$th VR user in the next time slot.


\subsubsection{Uplink Transmission} For conventional wireless VR system without FoV prediction, each VR user needs to deliver its actual request FoV to the MEC through uplink broadcast transmission. To focus on the rendering and downlink transmission, at the $t$th time slot, we assume that the received FoV $\widehat{F_{t}^{\text{oV}}}$ is equal to the actual requested FoV $F_{t}^{\text{oV}}$ following
\begin{equation}
    \widehat{F_{t}^{\text{oV}}} = F_{t}^{\text{oV}}.
\end{equation}
For our proposed system with FoV prediction, the uplink received FoV $F_{t}^{\text{oV}}$ after prediction will be used to check the correctness of predicted FoV $\widetilde{F_{t}^{\text{oV}}}$. We define the FoV prediction accuracy over $T$ time slots as 
\begin{equation}
    P_{\text{FoV}} = \frac{1}{T}\sum_{t=1}^{T}\left[\frac{F_{t}^{\text{oV}} - \widetilde{F_{t}^{\text{oV}}}}{F_{t}^{\text{oV}}}\right]\times 100\%.
\end{equation}

\begin{figure*}[ht]
    \centering
    \includegraphics[width=6.0in]{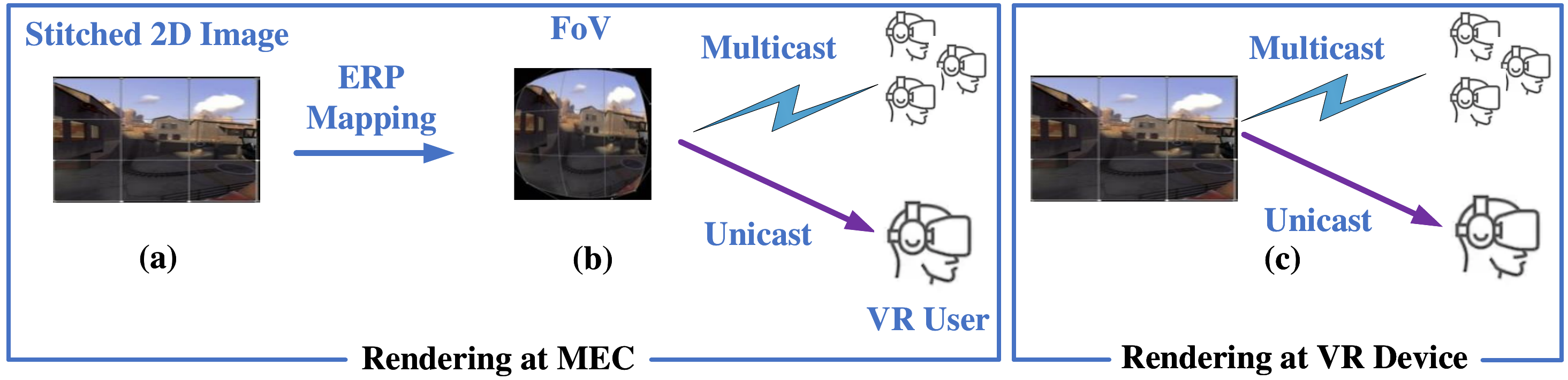}
    \caption{FoV rendering.}
    \label{basic_modules}
\end{figure*}

\subsubsection{FoV Rendering} When VR user requests to watch VR video frames, based on the predicted or uplink received FoV, the corresponding portion of the sphere can be rendered at MEC or VR device.
\begin{itemize}
    \item As shown in Fig. 3 (a) and (b), when the rendering function is executed at the MEC, a stitched 2D image, whose color model is RGB, will be rendered into the required FoV through equirectangular projection (ERP) mapping \cite{ERP} for downlink multicast or unicast transmission.
    \item When the rendering function is processed at the VR device, the stitched 2D picture frames will be first multicast or unicast to the VR users. Then, it will be rendered into the required FoV via ERP mapping, which is shown in Fig. 3 (c). 
    
\end{itemize}
Here, the number of pixels in the stitched 2D image is used to quantify the size of the executed data during FoV rendering. To evaluate the best rendering strategy,  we propose three FoV rendering schemes  as detailed below: 

(a) \textbf{MEC Rendering without Migration Scheme:} In this scheme, the rendering from stitched 2D image to 3D FoV occurs at each MEC with associated VR group.

(b) \textbf{MEC Rendering with Migration Scheme:} With different computational capabilities at each MEC, multiple VR user groups selecting the same FoV but different MECs only performs FoV rendering at only one MEC, and this selected MEC can migrate the rendered FoV to other MECs via fiber links to save the computational resources. 

(c) \textbf{VR Device Rendering Scheme:} This scheme is a conventional scheme for comparison, where the FoV rendering occurs only at the VR device, such that the stitched 2D picture frames need to be transmitted to the VR device for rendering locally using ERP mapping. Due to the fact that the computation ability of the VR device is much smaller than that of the MEC, we expect that it may cost much more time for the VR device to render the required FoV.

\subsubsection{Multi-group Multicast and Unicast} Based on the received FoV  in the uplink or the predicted FoV of each VR user, the VR users with the same received/predicted FoVs can be grouped together. After FoV rendering,  the MECs will multicast the required FoVs to VR users selecting the same FoV, or unicast a single VR user selecting unique FoV, respectively.
Let us consider a set of $\mathcal{B}$ =\{1, 2,..., B\} MECs, and each MEC is equipped with $N$ transmit antennas. These $B$ MECs serve the downlink transmission for $B$ VR user groups  $\mathcal{V} = \{\mathcal{V}_1, \mathcal{V}_2,..., \mathcal{V}_B\}$ with single antenna. The B VR user groups can be multicast group, unicast group, or inactive group with no VR users. Assuming that there are $M$ multicasting groups and $U$ unicasting groups ($M + U \leq B$), these $M$ muslticast groups and $U$ unicast groups can be denoted as the sets $\mathcal{V}^{\rm{mul}} = \{\mathcal{V}_{1}^{\rm{mul}}, \mathcal{V}_{2}^{\rm{mul}},...,\mathcal{V}_{M}^{\rm{mul}}$\} and $\mathcal{V}^{\rm{uni}} = \{\mathcal{V}_{1}^{\rm{uni}}, \mathcal{V}_{2}^{\rm{uni}},...,\mathcal{V}_{U}^{\rm{uni}}\}$, respectively. With the number of VR users in the $k$th multicasting group denoted as $|\mathcal{V}_{k}^{\rm{mul}}|$, the total number of VR users can be calculated as $K_{\text{VR}} = \sum_{k = 1}^{M}|\mathcal{V}_{k}^{\rm{mul}}| + U$. Note that each VR user can only be assigned to just one group.

\subsection{Mathematical Model}
\subsubsection{FoV Rendering Model} We denote the number of pixels as $\mathcal{R}$, and the size of each pixel is $8$ bits. For the MEC rendering schemes, at each time slot, the size of the FoV to be transmitted in the downlink can be calculated as
\begin{equation}
    C = \mathcal{R}\times \mathcal{R}\times 3\times 8\times V = 48\mathcal{R}^{2},
\end{equation}
where the single-eye resolution is $\mathcal{R}\times \mathcal{R}$, $3$ presents the red, green and blue color in RGB model, and $V$ is the number of viewpoints with $V=2$ for two eyes. According to \cite{Hu1}, the resolution of the FoV is at least $1080\text{p}$, and $C$ can be very large when $\mathcal{R}$ is high. Usually, the FoV has to be compressed  before downlink multicast or unicast. By assuming the compression ratio as $\mathcal{C_{\mathcal{R}}}$, the size of the compressed data for downlink transmission can be calculated as $\frac{C}{\mathcal{C}_{\mathcal{R}}}$. In addition, through ERP mapping, 25 percent pixels of the stitched 2D image can be reduced \cite{building_360}, which means that the data size of the FoV is 75 percent of that of the stitched 2D image. Thus, the data size of the stitched 2D image can be calculated as $\mathcal{M} = \frac{4}{3}C = 64\mathcal{R}^{2}$, and $64\mathcal{R}^{2}$ bits data are required to be executed in the ERP mapping step. 

\subsubsection{Downlink Transmission Model} For VR users in the multicast groups, the multicast signal between the $b$th MEC and the $k$th VR user in the $j$th multicast group at the $t$th time slot can be written as
\begin{align}
    \textbf{y}_{j_{k},b}^{\rm{mul}}(t) &= \textbf{h}_{j_{k},b}^{H}(t)\textbf{v}_{j,b}^{\rm{mul}}(t)x_{b}^{\rm{mul}}(t) + \\ \nonumber
    &\sum_{\mathcal{V}_{i}^{\rm{mul}}\in \mathcal{V}^{\rm{mul}}/\mathcal{V}_{b}^{\rm{mul}}, m\in\mathcal{V}_{i}^{\rm{mul}}}\textbf{h}_{j_{k},i}^{H}(t)\textbf{v}_{m,i}^{\rm{mul}}(t)x_{i}^{\rm{mul}}(t) + \\ \nonumber
    &\sum_{\mathcal{V}_{l}^{\rm{uni}}\in\mathcal{V}^{\rm{uni}}, u\in \mathcal{V}_{l}^{\rm{uni}}}\textbf{h}_{j_{k},l}^{H}(t)\textbf{v}_{u,l}^{\rm{uni}}(t)x_{u}^{\rm{uni}}(t) + 
    \textbf{n}_{j_{k}}(t) ,
\end{align}
where $\textbf{h}_{j_{k},b}(t)\in\mathbb{C}^{M\times 1} \sim \mathcal{CN}(\textbf{0}, \alpha\textbf{I}_{M})$ is the uncorrelated Rayleigh fading channel vector between the $b$th MEC and the $k$th VR user in the $j$th multicast group, $\alpha$ is the large-scale fading coefficient of the multiscast VR users. $\textbf{v}_{j,b}^{\rm{mul}}(t)\in{\mathbb{C}^{M\times 1}}$ and $\textbf{v}_{u,l}^{\rm{uni}}(t)\in{\mathbb{C}^{M\times 1}}$ are the multicast and unicast vectors from the $b$th and $l$th MECs connected to the VR users in the $j$th multicast group and the $u$th VR user in the unicast group, respectively. In (5), $x_{b}^{\rm{mul}}(t)$ and $x_{u}^{\rm{uni}}(t)$ are the multicast and unicast messages intended for the VR users in the $j$th multicast group and the $u$th VR user in the unicast group, respectively. We assume that $x_{b}^{\rm{mul}}(t)$ and $x_{u}^{\rm{uni}}(t)$ are independent from each other. Meanwhile, $\sum_{\mathcal{V}_{i}^{\rm{mul}}\in \mathcal{V}^{\rm{mul}}/\mathcal{V}_{b}^{\rm{mul}}, m\in\mathcal{V}_{i}^{\rm{mul}}}\textbf{h}_{j_{k},i}^{H}(t)\textbf{v}_{m,i}^{\rm{mul}}(t)x_{i}^{\rm{mul}}(t)$ and $\sum_{\mathcal{V}_{l}^{\rm{uni}}\in\mathcal{V}^{\rm{uni}}, u\in \mathcal{V}_{l}^{\rm{uni}}}\textbf{h}_{j_{k},l}^{H}(t)\textbf{v}_{u,l}^{\rm{uni}}(t)x_{u}^{\rm{uni}}(t)$ are the interference from the other MECs that provide FoVs for the VR users in other multicast and unicast groups, respectively. In addition, $\textbf{n}_{j_{k}}(t) \sim \mathcal{CN}(0, \sigma_{j_{k}}^{2}\textbf{I}_{M})$ is the additive white Gaussian noise at the $k$th VR user in the $j$th multicast group.

Based on (5), the multicast transmission rate between the $k$th VR user in the $j$th multicast group and the $b$th MEC at the $t$th time slot can be expressed as
\begin{equation}
    R_{j_{k},b}^{\rm{mul}}(t) = \log_{2}\left(1 + \frac{|\textbf{h}_{j_{k},b}^{H}(t)\textbf{v}_{j,b}^{\rm{mul}}(t)|^2}{ \textbf{I}_{j_{k},b}^{\rm{mul}}(t) +  \sigma_{j_{k}}^{2}}\right),
\end{equation}
where
\begin{equation}
    \textbf{I}_{j_{k},b}^{\rm{mul}}(t) = \!\!\!\!\!\!\!\!\!\!\!\!\sum\limits_{\begin{subarray}{l}\mathcal{V}_{i}^{\rm{mul}}\in\mathcal{V}^{\rm{mul}}/\mathcal{V}_{b}^{\rm{mul}}\\ m\in\mathcal{V}_{i}^{\rm{mul}} \end{subarray}}\!\!\!\!\!\!\!\!\!\!\!\!\!\!|\textbf{h}_{j_{k},i}^{H}(t)\textbf{v}_{m,i}^{\rm{mul}}(t)|^2 \!+\! \!\!\!\!\!\!\!\!\sum\limits_{\begin{subarray}{l}\mathcal{V}_{l}^{\rm{uni}}\in\mathcal{V}^{\rm{uni}}\\u\in \mathcal{V}_{l}^{\rm{uni}}\end{subarray}}\!\!\!\!\!\!|\textbf{h}_{j_{k},l}^{H}(t)\textbf{v}_{u,l}^{\rm{uni}}(t)|^2.
\end{equation}

For the VR users in the unicast groups, the unicast signal between the $b$th MEC and the $k$th VR user at the $t$th time slot can be expressed as
\begin{align}
   \textbf{y}_{k,b}^{\rm{uni}}(t) &= \textbf{g}_{k,b}^{H}(t)\textbf{v}_{k,b}^{\rm{uni}}(t)x_{b}^{\rm{uni}}(t) + \\ \nonumber
    &\sum_{\mathcal{V}_{i}^{\rm{uni}}\in\mathcal{V}^{\rm{uni}}/\mathcal{V}_{b}^{\rm{uni}},u\in \mathcal{V}_{i}^{\rm{uni}}}\textbf{g}_{k, i}^{H}(t)\textbf{v}_{u,i}^{\rm{uni}}(t)x_{i}^{\rm{uni}}(t) + \\ \nonumber
    &\sum_{\mathcal{V}_{l}^{\rm{mul}}\in \mathcal{V}^{\rm{mul}}, m\in\mathcal{V}_{l}^{\rm{mul}}}\textbf{g}_{k,l}^{H}(t)\textbf{v}_{m,l}^{\rm{mul}}(t){x}_{l}^{\rm{mul}}(t) + 
    \textbf{n}_{k}(t),
\end{align}
where $\textbf{g}_{k,b}(t)\in\mathbb{C}^{M\times 1} \sim\mathcal{CN}(\textbf{0}, \beta\textbf{I}_{M})$ is the uncorrelated Rayleigh fading channel vector between the $b$th MEC and the $k$th VR user in the unicast group, and $\beta$ is the large-scale fading coefficient for the unicast VR users. Meanwhile, $\sum_{\mathcal{V}_{i}^{\rm{uni}}\in\mathcal{V}^{\rm{uni}}/\mathcal{V}_{b}^{\rm{uni}}, u\in\mathcal{V}_{i}^{\rm{uni}}}\textbf{g}_{k,i}^{H}(t)\textbf{v}_{u,i}^{\rm{uni}}(t)x_{i}^{\rm{uni}}(t)$ and $\sum_{\mathcal{V}_{l}^{\rm{mul}}\in \mathcal{V}^{\rm{mul}},m\in\mathcal{V}_{l}^{\rm{mul}}}\textbf{g}_{k,l}^{H}(t)\textbf{v}_{m,l}^{\rm{mul}}(t){x}_{l}^{\rm{mul}}(t)$ are the interference from the other MECs that provide service for VR users in the unicast and multicast groups, respectively. In (8), $\textbf{n}_k(t) \sim \mathcal{CN}(0, \sigma_{k}^{2}\textbf{I}_{M})$ is the additive white Gaussian noise at the $k$th VR user in the unicast group.

Based on (8), the unicast transmission rate between the $k$th VR user and the $b$th MEC at the $t$th time slot can be written as
\begin{equation}
    R_{k,b}^{\rm{uni}}(t) = \log_{2}\left(1 + \frac{|\textbf{g}_{k,b}^{H}(t)\textbf{v}_{k,b}^{\rm{uni}}(t)|^2}{\textbf{I}_{k,b}^{\rm{uni}}(t) + \sigma_{k}^{2}}\right),
\end{equation}
where
\begin{equation}
    \textbf{I}_{k,b}^{\rm{uni}}(t) = \!\!\!\!\!\!\!\!\!\!\sum\limits_{\begin{subarray}{l}\mathcal{V}_{i}^{\rm{uni}}\in\mathcal{V}^{\rm{uni}}/\mathcal{V}_{b}^{\rm{uni}}\\u\in \mathcal{V}_{i}^{\rm{uni}}\end{subarray}}\!\!\!\!\!\!\!|\textbf{g}_{k, i}^{H}(t)\textbf{v}_{u,i}^{\rm{uni}}(t)|^2 \!\!+\!\!\!\!\!\! \!\!\!\!\sum\limits_{\begin{subarray}{l}\mathcal{V}_{l}^{\rm{mul}}\in \mathcal{V}^{\rm{mul}}\\ m\in\mathcal{V}_{l}^{\rm{mul}}\end{subarray}}\!\!\!\!\!\!\!|\textbf{g}_{k,l}^{H}(t)\textbf{v}_{m,l}^{\rm{mul}}(t)|^2.
\end{equation}

\begin{figure*}[ht]
    \centering
    \includegraphics[width=6 in]{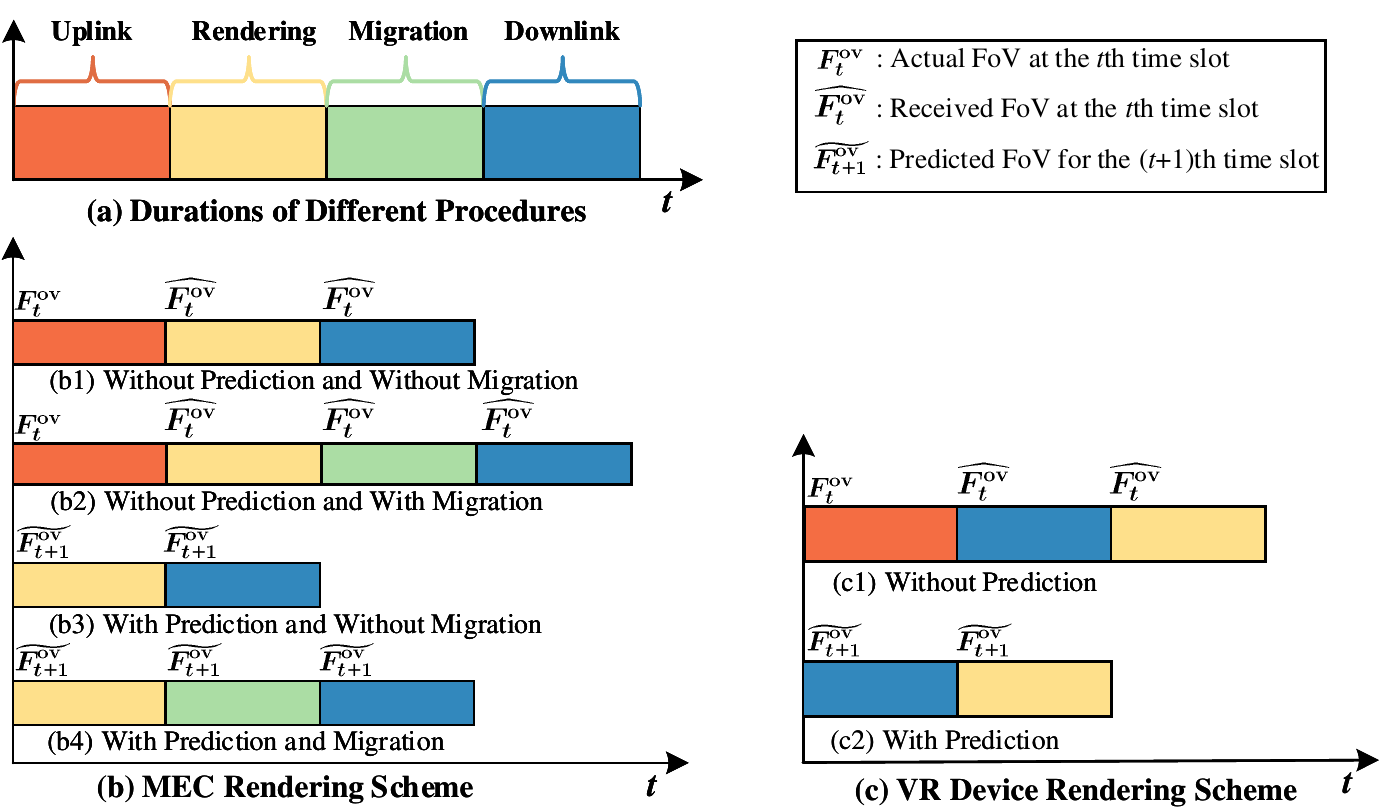}
    \caption{VR interaction latency of the proposed MEC and VR device rendering schemes.}
    \label{basic_modules}
\end{figure*}

\subsubsection{VR Interaction Latency} As defined in \cite{Hu1,XR}, the VR interaction latency $T^{\text{loop}}$ is the time starting from the VR user's movement to the time where the virtual environment responds to its movements. It is consisted of four parts: 1) the time of the VR user to uplink its FoV request, and other tracking information ($T^{\text{uplink}}$); 2) the time of the FoV rendering at MECs or VR devices to generate the predicted or uplink requested FoV ($T^{\text{render}}$); 3) the time to migrate the rendered FoV from one optimal MEC to the other MECs ($T^{\text{migration}}$) with the VR groups selecting the same FoV via fiber; and 4) the time to transmit the rendered FoV or the stitched 2D picture frames from the MEC to the VR user ($T^{\text{downlink}}$) depending on rendering at the MEC or the VR device, respectively. Thus, the VR interaction latency $T^{\text{loop}}$ can be calculated as
\begin{equation}
    T^{\text{loop}} = T^{\text{uplink}} + T^{\text{render}} + T^{\text{migration}}+ T^{\text{downlink}},
\end{equation}
as shown in Fig. 4 (a).

Let us assume the execution ability of the GPU of the $k$th MEC or VR device as $F_{k}^{\text{MEC}}$ and $F_{k}^{\text{VR}}$, respectively. We use $f_{k}^{\text{MEC}}$ and $f_{k}^{\text{VR}}$ to represent the number of cycles required for processing one bit of input data of the $k$th MEC or VR device, respectively. Here, the number of cycles depends
on the application type and the GPU architecture of the $k$th MEC or VR device. For the MEC rendering scheme with migration, we assume that the $b$th MEC is selected to be only rendering MEC with the same FoV request, the distance between the $k$th MEC and the $b$th MEC is $\hat{L}_{k,b}$, and the trasnmission rate of the optical fiber is $R^{\text{fiber}}$. 

The VR interaction latency of the proposed MEC and VR device rendering schemes are shown in Fig. 4 (b) and (c) and introduced in details as follows:

(a) \textbf{MEC Rendering without Migration:} 
\textit{a-1) Without prediction:} At the $t$th time slot, the VR user needs to deliver the actual FoV request of VR users $F_{t}^{\text{oV}}$ through uplink broadcast transmission, and the rendering of received FoV $\widehat{F_{t}^{\text{oV}}}$ is executed at each MEC with associated VR user. As shown in Fig. 4 (b1), at the $t$th time slot, if the $k$th VR user is served by the $k$th MEC, the VR interaction latency can be calculated as
\begin{align}
    T_{k}^{\text{loop}} &= T_{k}^{\text{uplink}} + T_{k}^{\text{render}}+ T_{k}^{\text{downlink}}\\ \nonumber
    &=T_{k}^{\text{uplink}} + \frac{f_{k}^{\text{MEC}}\mathcal{M}}{F_{k}^{\text{MEC}}} + \frac{C}{\mathcal{C}_{\mathcal{R}}R_{k,k}^{\text{down}}},
\end{align}
where $R_{k,k}^{\text{down}}\in\{R_{k,k}^{\rm{mul}}, R_{k,k}^{\rm{uni}}\}$, and $R_{k,k}^{\rm{mul}}$ and $R_{k,k}^{\rm{uni}}$ are given in (6) and (9).

\textit{a-2) With prediction:} As shown in Fig. 4 (b3), based on the uplink received FoVs at the $t$th time slot $\widehat{F_{t}^{\text{oV}}}$ and several previous time slots, we predict the FoV preference of each VR user  at the $(t+1)$th time slot $\widetilde{F_{t+1}^{\text{oV}}}$. For the $k$th VR user directly served by the $k$th MEC, the VR interaction latency with predicted FoV can be written as (12) with $T_{k}^{\text{uplink}} = 0$.

(b) \textbf{MEC Rendering with Migration:} \textit{b-1) Without prediction:} For the VR user groups requesting the same FoV $\widehat{F_{t}^{\text{oV}}}$ through uplink transmission only select one MEC for rendering, and the rendered FoV can be migrated to other MECs. As shown in Fig. 4 (b2), at the $t$th time slot, if the $k$th VR user directly served by the $b$th MEC performs rendering itself, the VR interaction latency of the required FoV of the $k$th VR user can be presented as 
\begin{align}
    T_{k}^{\text{loop}} &= T_{k}^{\text{uplink}} + T_{k}^{\text{render}}+ T_{k}^{\text{downlink}}\\ \nonumber &= T_{k}^{\text{uplink}} + \frac{f_{b}^{\text{MEC}}\mathcal{M}}{F_{b}^{\text{MEC}}} + \frac{C}{\mathcal{C}_{\mathcal{R}}R_{k,b}^{\text{down}}},
\end{align}
where $R_{k,b}^{\text{down}} = R_{k,b}^{\rm{mul}}$ and $R_{k,b}^{\rm{mul}}$ is given in (6).

If the $k$th VR user is served by the $k$th MEC, where the rendering is not performed by itself, but by the $b$th MEC, the interaction latency of the $k$th VR user can be presented as
\begin{align}
    T_{k}^{\text{loop}} &= T_{k}^{\text{uplink}} + T_{k}^{\text{render}} + T_{k}^{\text{migration}} + T_{k}^{\text{downlink}} \\ \nonumber &= T_{k}^{\text{uplink}} + \frac{f_{b}^{\text{MEC}}\mathcal{M}}{F_{b}^{\text{MEC}}} + \frac{\hat{L}_{k,b}}{R^{\text{fiber}}} + \frac{C}{\mathcal{C}_{\mathcal{R}}R_{k,k}^{\text{down}}},
\end{align}
where $R_{k,k}^{\text{down}} = R_{k,k}^{\rm{mul}}$ and $R_{k,k}^{\rm{mul}}$ is given in (6).


\textit{b-2) With prediction:} As shown in Fig. 4 (b4), at the $t$th time slot, for the $k$th VR user directly served by the $b$th MEC, the VR interaction latency with the predicted FoV of the $k$th VR user can be denoted as (13) with $T_{k}^{\text{uplink}} = 0$.

Otherwise, the predicted FoV will be migrated to the $k$th MEC from the $b$th MEC, and the VR interaction latency of the predicted FoV of the $k$th VR user can be denoted as (14) with $T_{k}^{\text{uplink}} = 0$.

(c) \textbf{VR Device Rendering:} \textit{c-1) Without prediction:} According to Fig. 4 (c1), at the $t$th time slot, for the $k$th VR user served by the $k$th MEC without FoV prediction, the VR interaction latency of the $k$th VR user can be written as
\begin{align}
    T_{k}^{\text{loop}} &= T_{k}^{\text{uplink}} + T_{k}^{\text{downlink}}+ T_{k}^{\text{render}}\\ \nonumber &= T_{k}^{\text{uplink}} + \frac{\mathcal{M}}{\mathcal{C}_{\mathcal{R}}R_{k,k}^{\text{down}}} + \frac{f_{k}^{\text{VR}}\mathcal{M}}{F_{k}^{\text{VR}}},
\end{align}
where $R_{k,k}^{\text{down}}\in\{R_{k,k}^{\rm{mul}}, R_{k,k}^{\rm{uni}}\}$, and $R_{k,k}^{\rm{mul}}$ and $R_{k,k}^{\rm{uni}}$ are given in (6) and (9).

\textit{c-2) With prediction:} As shown in Fig. 4 (c2), at the $t$th time slot, the FoV preference of the VR users for the $(t+1)$th time slot will be predicted, and the VR interaction latency with the predicted FoV of the $k$th VR user served by the $k$th MEC can be presented as (15) with $T_{k}^{\text{uplink}} = 0$.

\subsubsection{VR Quality of Experience} The quality of the FoV can be influenced by many factors, such as blockiness, blur, contrast distortion, freezing, colour depth, sharpness, etc \cite{viprovoq,8293688}. To evaluate the performance of the proposed MEC rendering schemes, we focus on the objective in maximizing the Peak Signal-to-Noise Ratio (PSNR) \cite{TechniqueQoE}, knowing that it is the most common and simple objective VR video quality assessment, and the PSNR is usually defined by the Mean Squared Error (MSE)  of the $k$th VR user between an initial FoV $\mathcal{I}_k$ and the distorted FoV $\mathcal{D}_k$. According to \cite{QoEmodel}, to  measure the QoE of the $k$th VR user based on the MSE, we propose a binary function where $\mathcal{I}_k = 1$ and $\mathcal{D}_k \in\{0, 1\}$ to represent whether the FoV can be rendered and delivered within the threshold of VR interaction latency of the $k$th VR user. For real-time interactive VR applications, the delayed FoV will bring unpleasant human experience, thus, for the $k$th VR user, we revise basic QoE model by incorporating a maximum VR interaction latency requirement $T_k^{\text{th}}$. More specifically, if  $T_{k} \leq T_k^{\text{th}}$, the rendered FoV is regarded as successfully delivered to VR device, then $\mathcal{D}_k = 1$, otherwise, $\mathcal{D}_k = 0$. The MSE of the $k$th VR user can be written as 
\begin{equation}
    \rm{MSE}_{k} = (\mathcal{I}_k - \mathcal{D}_k)^2.
\end{equation}
According to \cite[Eq. (2)]{QoEmodel}, the $\rm{PSNR}$ of the $k$th VR user is defined as 
\begin{equation}
    \rm{PSNR}_{k} = 10\log_{10}\frac{1}{\rm{MSE}_{k}}.
\end{equation}
As can be seen from (17), for $\rm{MSE}_{k} = 0$, $\rm{PSNR}_{k}\rightarrow\infty$. To avoid the infinite value of $\rm{PSNR}$, we introduce a positive number $\bigtriangleup$ and modify (17) as
\begin{equation}
    \rm{PSNR}_{k} = 10\log_{10}\frac{1 + \bigtriangleup}{\rm{MSE}_{k} + \bigtriangleup},
\end{equation}
where $\bigtriangleup > 0$ and we set $\bigtriangleup = 1$ in this paper.

\subsection{Problem Formulation}
To ensure that each requested FoV is rendered and transmitted within the VR interaction latency, we aim to optimize the total QoE under fixed VR interaction latency constraint via determining the optimal association between MEC and VR user group, and  optimal rendering MEC for model migration. 

The proposed MEC rendering schemes aim at maximizing the long-term total QoE under VR interaction latency constraint in the continuous time slots with respect to the policy $\pi$ that maps the current state information $S_t$ to the probabilities of selecting possible actions in $A_t$. Therefore, based on the QoE of each VR user, an optimization problem (P1) is formulated as
\begin{align}
    (\rm{P1})~\max_{\pi(A_t|S_t)}&\sum_{t=0}^{\infty}\sum_{k = 1}^{K_{\text{VR}}}\gamma^{t}\mathbb{E}_{\pi}[\rm{PSNR}_{k}]\\
    &T_{k} \leq T_k^{\text{th}},
\end{align}
where $\gamma\in[0, 1)$ is the discount factor which can determine the weight of the future QoE, and $\gamma = 0$ means that the agent just concerns the immediate reward. The state $S_t$ contains the index of the requested FoV, the location of each VR user, and the computation ability of each MEC. The action $A_{t}$ includes the optimal association between MEC and VR user group, and optimal rendering MEC for model migration.

Since the dynamics of the wireless VR system is Markovian in continuous time slots, this is a Partially Observable Markov Decision Process (POMDP) problem which is generally intractable. Here, the parital observation refers to that the MECs can only know the previous FoV requests and the location of each VR user in the environment, while they are unable to know all the information of the communication environment, including, but not limited to, the channel conditions, and the FoV request in the current time slot.  Furthermore, the traditional optimization methods may need the global information to achieve the optimal solution, which not only increase the overhead of signal transmission, but also increase the computation complexity. Approximate solutions will be discussed in Section III.

\begin{figure*}[!t]
    \centering
    \includegraphics[width=7 in]{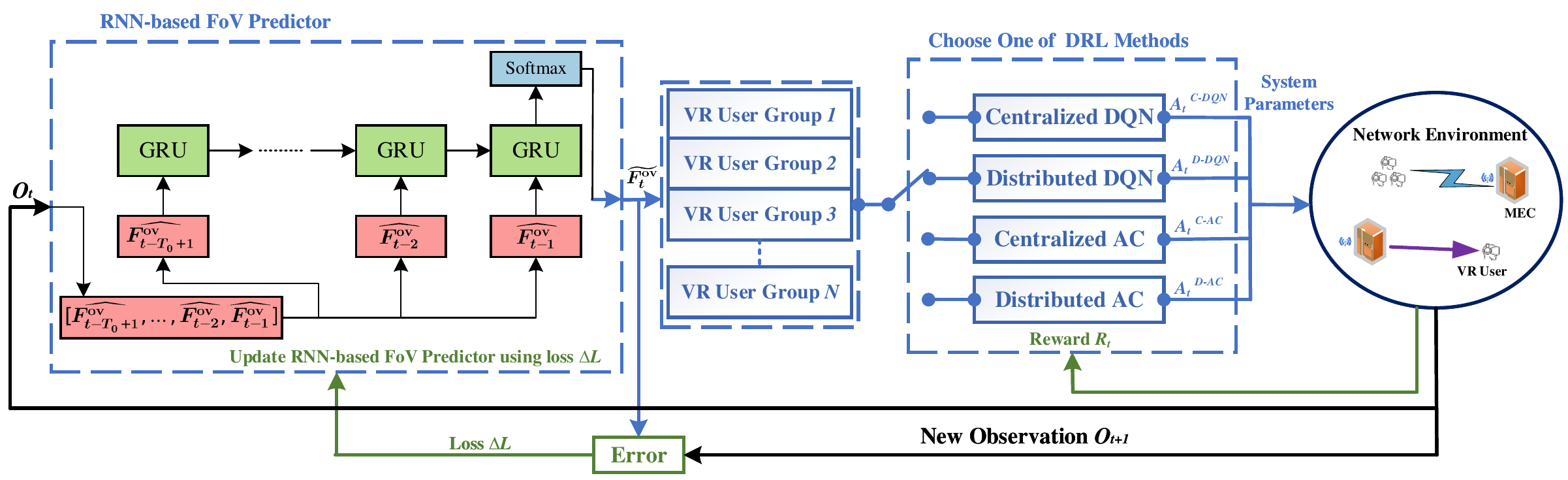}
    \caption{Decoupled learning strategy for MEC rendering schemes in the wireless VR network.}
    \label{basic_modules}
\end{figure*}

\section{Deep Reinforcement Learning-Based MEC Rendering Scheme}
Knowing the deep neural networks as one of the most impressive non-linear approximation functions, DRL is an effective method to optimally solve POMDP problems \cite{Jiang2}. In this section, to solve (P1), a decoupled learning strategy is proposed for FoV prediction and MEC rendering association, as shown in Fig. 5. Specifically, a RNN model based on GRU is used to predict FoV preference of each user over time. Then, four DRL algorithms, including centralized DQN, distributed DQN, centralized AC, and distributed AC, are proposed to select the FoV rendering MEC and the associated MEC for downlink transmission.


\subsection{FoV Prediction}
Brownian Motion is used to simulate the eye movement of VR users over time, and assuming that the uplink received FoV preference of the $k$th VR user at the $t$th time slot is $\widehat{{F}_{t}^{\text{oV}}}^{k}\in\{1,2,...,N_{\text{FoV}}\}$. In order to detect dynamics in FoV preference of each VR user, the proposed learning scheme aims at utilizing not only the information presents in the most recent observation $O_{t}=\{O_t^{1},O_t^{2},...,O_t^{\text{K}_{\text{VR}}}\}$, where $O_t^{k} = \{\widehat{{F}_{t}^{\text{oV}}}^{k}\}$, but also the historical information in the previous observations ${H}_{t}=\{O_{t-T_{0}+1},...,O_{t-2},O_{t-1}\}$ given a memory window $T_{0}$. To recognize FoV preference over time, a RNN model with parameters $\pmb{\theta}_{\text{RNN}}$, and specifically a GRU architecture, is leveraged. $\pmb{\theta}_{\text{RNN}}$ is consisted of both the GRU internal parameters and the weights of the softmax layer. We choose RNN due to its ability in capturing time correlation of FoV preference over time, which can help learn the time-varying FoV preference for better prediction accuracy.

As shown in Fig. 5, the GRU layer includes multiple standard GRU units and historical observations $[O_{t-T_{0}+1},...,O_{t-1},]$ sequentially inputted into the RNN predictor. For the $k$th VR user, the GRU layer is connected to an output layer which is consisted of a softmax non-linearity with ${N}_{\text{FoV}}$ output values, which represents the predicted probability $\mathcal{P}\{\widehat{{F}_{t}^{\text{oV}}}^{k} = \hat{f}|[O_{t-T_{0}+1}^{k},...,O_{t-1}^{k}], \pmb{\theta}_{\text{RNN}}\}$ of the $\hat{f}$th FoV ($\hat{f} = 1,...,{N}_{\text{FoV}}$) for the $t$th time slot given historical observations $[O_{t-T_{0}+1}^{k},...,O_{t-1}^{k}]$. 

To adapt the model parameter $\pmb{\theta}_{\text{RNN}}$, standard Stochastic Gradient Descent (SGD) via BackPropagation Through Time (BPTT) \cite{BPTT} is deployed. At the $(t+1)$th time slot, the parameters $\pmb{\theta}_{\text{RNN}}$ of the RNN predictor can be updated as
\begin{equation}
    \pmb{\theta}_{\text{RNN}}^{t+1} = \pmb{\theta}_{\text{RNN}}^{t} - \lambda_{\text{RNN}}\nabla L_{\text{RNN}}(\pmb{\theta}_{\text{RNN}}^{t}),
\end{equation}
where $\lambda_{\text{RNN}}\in(0,1]$ is the learning rate, $\nabla L_{\text{RNN}}(\pmb{\theta}_{\text{RNN}}^{t})$ is the gradient of the loss function $L_{\text{RNN}}(\pmb{\theta}_{\text{RNN}}^{t})$ to train the RNN predictor. $L_{\text{RNN}}(\pmb{\theta}_{\text{RNN}}^{t})$ can be obtained by averaging the cross-entropy loss as 
\begin{equation}
    L_{\text{RNN}}^{t}(\pmb{\theta}_{\text{RNN}}) \!=\! -\!\!\!\!\!\!\!\!\sum_{t^{'} = t-T_b+1}^{t}\!\!\!\!\!\log\left(\!\mathcal{P}\{\widehat{{F}^{\text{oV}}_{t^{'}}} = \widetilde{{F}^{\text{oV}}_{t^{'}}}|O_{t^{'}-T_{0}}^{t^{'}},\pmb{\theta}_{\text{RNN}}\}\!\right),
\end{equation}
where
\begin{equation}
    O_{t^{'}-T_{0}}^{t^{'}} = [O_{t^{'}-T_{0}+1},...,O_{t^{'}-1},O_{t^{'}}],
\end{equation}
and $T_b$ is the randomly selected mini-batch size.

Through FoV prediction, MECs are able to know the FoV preference of each VR user in advance. The VR users with the same predicted FoVs can be grouped together. After FoV rendering, the MECs will multicast or unicast the required FoVs to VR users selecting the same FoV, or a single VR user selecting unique FoV, respectively.

\subsection{Deep Reinforcement Learning}
The main purpose of Reinforcement Learning (RL) is to select proper MECs for MEC rendering schemes. Through a series of action strategies, MECs are able to interact with the environment, and obtain rewards due to their actions, which help to improve their action strategies. After plenty of iterations, MECs can learn the optimal policy that maximizes the long-term rewards. 

We define $S\in\mathcal{S}$, $A\in\mathcal{A}$, and $R\in\mathcal{R}_{e}$ as any state, action and reward from their corresponding sets, respectively. According to the observed environmental state $S_t$ at the $t$th time slot, MECs choose specific actions $A_t$ from the set $\mathcal{A}$ and receive rewards $R_t$, which are regarded as a metric to measure whether the selected actions are good. Thus, the purpose of RL algorithm is to find an optimal policy $\pi$ which can maximize the long-term reward for $A = \pi(S)$. The optimization function can be formulated as $<S, A, R>$ and the detailed descriptions of the state, action and reward of problem (P1) are introduced as follows.
\begin{itemize}
    \item State: At the $t$th time slot, the network state can be denoted as 
    \begin{align}
        S_{t} &= (\widetilde{\mathcal{F}_{t}^{\text{oV}}}, \mathcal{L}_{k,i}^{t}, \mathcal{F}_{i}^{\text{MEC}})\in \mathcal{S},  \\ \nonumber
        \text{with}~\widetilde{\mathcal{F}_{t}^{\text{oV}}} &= \{\widetilde{{F}_{t}^{\text{oV}}}^{1}, \widetilde{{F}_{t}^{\text{oV}}}^{2},...,\widetilde{{F}_{t}^{\text{oV}}}^{K_{\text{VR}}}\}, \\ \nonumber
        \mathcal{L}_{k,i}^{t} &= \{ {l}_{k,1}^{t}, {l}_{k,2}^{t},...,  {l}_{k,B}^{t}, \},\\ \nonumber
        \mathcal{F}_{i}^{\text{MEC}} &= \{F_{1}^{\text{MEC}}, F_{2}^{\text{MEC}},..., F_{B}^{\text{MEC}}\},
    \end{align}
    where $\widetilde{{F}_{t}^{\text{oV}}}^{k}$ is the index of the predicted FoV of the $k$th VR user at the $t$th time slot. $l_{k,i}^{t}$ is the distance between the $k$th VR user and the $i$th MEC at the $t$th time slot. $F_{i}^{\text{MEC}}$ is the computation capability of the $i$th MEC.
    
    \item Action: The action space can be written as
    \begin{align}
        {A}_{t} &= \{\check{\mathcal{A}}_{k,q}^{t}, \acute{\mathcal{A}}_{k,i}^{t}\}\in\mathcal{A}, \\ \nonumber
        \text{with}~\check{\mathcal{A}}_{k,q}^{t} &= \{\check{A}_{k,1}, \check{A}_{k,2},...,\check{A}_{k,{N}_{\text{FoV}}}\},\\ \nonumber
        \acute{\mathcal{A}}_{k,i}^{t} &= \{\acute{A}_{k,1}, \acute{A}_{k,2},...,\acute{A}_{k,K_{\text{VR}}}\}, \nonumber
    \end{align}
    where $\check{A}_{k,q}^{t}\in\{0, 1\}$ and $\acute{A}_{k,i}^{t}\in\{0, 1\}$ represent whether the $k$th MEC will render the $q$th FoV and serve the $i$th VR user at the $t$th time slot, respectively. For instance, if $\check{A}_{k,q}^{t} = 1$ and $\check{A}_{k,j}^{t} = 0$ , the $k$th MEC will render and migrate the $q$th FoV to the $j$th ($j\neq k$) MEC choosing the same FoV. If $\acute{A}_{k,i}^{t} = 1$, the $k$th MEC will support the downlink transmission of the $i$th VR user, otherwise, not.
    
    \item Reward: The immediate reward $R_t$ is designed as
    \begin{equation}
        {R}_{t}(S_t,A_t) = \sum\limits_{k=1}^{K_{\text{VR}}} \text{PSNR}_{k}^{t}.
    \end{equation}
\end{itemize}

Thus, the discounted accumulation of the long-term reward can be denoted as
\begin{equation}
    {V}(S,\pi) = \sum_{t = 1}^{\infty}(\gamma)^{t - 1}{R}_{t}(S_t,A_t),
\end{equation}
where $\gamma\in[0,1)$ is the discount factor.

When the number of MECs and VR users are small, RL algorithm can efficiently obtain the optimal policy. However, when a large number of MECs and VR users exist, the state and action spaces will be scaled proportionally, which will inevitably result in massive computation latency and severely affect the performance of the RL algorithm. To address this issue, deep learning is introducted to RL, namely, deep reinforcement learning (DRL), through interaction with the environment, DRL can directly control the behavior of each agent, and solve complex decision-making problem. In DRL algorithm, two methods can be used to obtain optimal policy. One is called
value-based optimization, such as DQN, which indirectly optimizes the policy by optimizing value function. While the other is policy-based optimization, such as AC, which can directly optimize the policy. In the following sections, four DRL algorithms are introduced in detail.

\begin{figure}[ht]
    \centering
    \includegraphics[width=3.0 in]{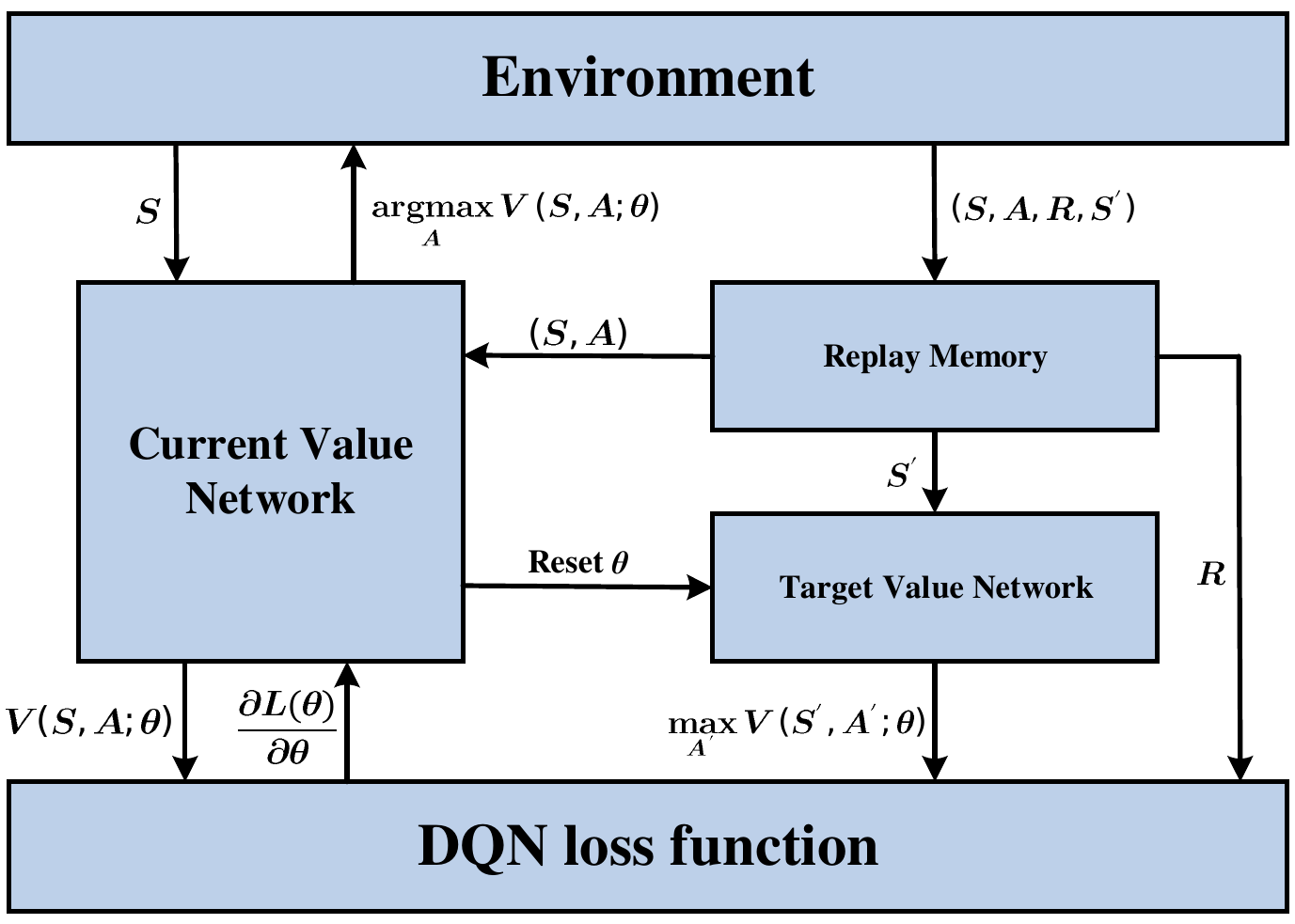}
    \caption{The DQN diagram of the MEC rendering scheme.}
    \label{basic_modules}
\end{figure}

\subsubsection{Centralized DQN}

As a value-based DRL algorithm, DQN combines a neural network with Q-learning and approximates the state-action value function via the deep neural network (DNN). Using DQN algorithm, a fraction of states are sampled and the neural network is applied to train a sufficiently accurate state-action value function, which is able to effectively solve the problem of high dimensionality in state space. Furthermore, the DQN algorithm uses the experience replay to train the learning process of RL. When updating the DQN algorithm, some experiences in the experience replay will be selected randomly to learn, so that the correlation among the training samples can be broke and the efficiency of the neural network can be improved. In addition, through averaging the selected samples, the distribution of training samples can be smoothed, which avoids the training divergence.

As shown in Fig. 6, the action-state value function $V_{\text{DQN}}(S,A)$ in the DQN agent can be parameterized by using a function ${V}_{\text{DQN}}(S,A;\pmb{\theta}_{\text{DQN}})$, where $\pmb{\theta}_{\text{DQN}}$ is the weight matrix of the DNN with multiple layers. Consider the conventional DNN, where the neurons between two adjacent layers are fully connected, which is so-called fully-connected layers. The input of the DNN is the variables in state $S_t$; the hidden layers are Rectifier Linear Units (ReLUs) through utilizing the function $f(x) = \max(0,x)$; the output layer is consisted of linear units, which are all available actions in $A_t$. The exploitation is obtained by performing propagation of ${V}_{\text{DQN}}(S,A;\pmb{\theta}_{\text{DQN}})$ with respect to the observed state $S_t$. Moreover, the parameter $\pmb{\theta}_{\text{DQN}}$ can be updated by using SGD as
\begin{equation}
    \pmb{\theta}_{\text{DQN}}^{t+1} = \pmb{\theta}_{\text{DQN}}^{t} - \lambda_{\text{DQN}}\nabla L_{\text{DQN}}(\pmb{\theta}_{\text{DQN}}^{t}),
\end{equation}
where $\lambda_{\text{DQN}}\in(0,1]$ is the learning rate, $\nabla L_{\text{DQN}}(\pmb{\theta}_{\text{DQN}}^{t})$ is the gradient of the loss function $L_{\text{DQN}}(\pmb{\theta}_{\text{DQN}}^{t})$ utilized to train the state-action value function. The loss function can be defined as
\begin{equation}
    L_{\text{DQN}}(\pmb{\theta}_{\text{DQN}}^{t}) =(\hat{V}_{\text{DQN}}-{V}_{\text{DQN}}(S_i,A_i;\pmb{\theta}_{\text{DQN}}^{t}))^{2},
\end{equation}
where
\begin{equation}
    \hat{V}_{\text{DQN}} = R_{i+1}+\gamma\max_{A} {V}_{\text{DQN}}(S_{i+1}, A; \Bar{\pmb{\theta}}_{\text{DQN}}^{t}).
\end{equation}
$(S_i,A_i,S_{i+1},R_{i+1})$ are randomly selected previous samples for some $i\in\{t-M_r,...,t\}$ with respect to a so-called minibatch. $M_r$ is the replay memory. $\Bar{\pmb{\theta}}_{\text{DQN}}^{t}$ is the so-called target Q-network which is utilized to estimate the future value of the Q-function in the update rule. Meanwhile, $\Bar{\pmb{\theta}}_{\text{DQN}}^{t}$ is periodically copied from the current value $\pmb{\theta}_{\text{DQN}}^{t}$ and kept fixed for some episodes. The use of minibatch, rather than a single sample, to update the state-action value function ${V}_{\text{DQN}}(S,A;\pmb{\theta}_{\text{DQN}})$ is able to improve the convergent reliability of value function.

Through deriving the loss function in (29) and calculating the expectation of the selected previous samples in minibatch, ${V}_{\text{DQN}}^{*}(S,A)$ can be obtained. The DQN algorithm is presented in Algorithm 1.

\begin{algorithm}[t]
\begin{algorithmic}[1]
\caption{DQN to dynamic decision-making and optimization of the MEC rendering scheme}
\STATE Initialize replay memory $D$ to capacity $\hat{\mathcal{N}}$, learning rate $\lambda_{\text{DQN}}\in(0,1]$ and discount factor $\gamma\in[0,1)$.
\STATE Initialize state-action value function ${V}_{\text{DQN}}(S,A;\pmb{\theta}_{\text{DQN}})$, the parameters of primary Q-network $\pmb{\theta}_{\text{DQN}}$ and target Q-network $\Bar{\pmb{\theta}}_{\text{DQN}}$.
\FOR{episode = 1,...,$M$}
    \STATE Input the network state $S$ of the MEC rendering scheme.
    \FOR{t = 1,...,T}
        \STATE Use $\epsilon$-greedy algorithm to select a random action $A_t$ from action space $\mathcal{A}$.
        \STATE Otherwise, select $A_t = \max\limits_{A\in \mathcal{A}}{V}(S_t,A;\pmb{\theta}_{\text{DQN}})$.
        \STATE The selected MECs render the predicted or uplink received FoVs and multicast/unicast them to VR users according to the selected action $A_t$.
        \STATE MECs observe reward $R_t$ and new state $S_{t+1}$.
        \STATE Store transition $(S_{t},A_{t},R_{t},S_{t+1})$ in replay memory $D$.
        \STATE Sample random minibatch of transitions $(S_j,A_j,R_j,S_{j+1})$ from replay memory $D$.
        \IF{$j+1$ is terminal}
            \STATE $y_j^{target} = R_j$.
        \ELSE
            \STATE $y_j^{target} = R_{j+1} + \gamma\max\limits_{A}{V}_{\text{DQN}}(S_{j+1},A;\pmb{\theta}_{\text{DQN}})$.
        \ENDIF
        \STATE Perform a gradient descent step and update parameters $\pmb{\theta}_{\text{DQN}}$ according to (28).
        \STATE Update parameter $\Bar{\pmb{\theta}}_{\text{DQN}}$ of the target network every $\Bar{K}$ steps.
    \ENDFOR
\ENDFOR
\end{algorithmic}
\end{algorithm}

\subsubsection{Distributed DQN}
In the centralized DRL algorithm, it learns a single optimization policy centrally at the central controller, which requires the global observations, rewards, and actions of each MEC. When the number of MECs and VR users increase, the size of the proposed model and parameters can expand exponentially. In this case, the GPU memory in central controller not only needs to hold the model and batch of data, but also the intermediate outputs of the feedforward computation. With dense VR users, GPU memory can be easily overloaded in practice, especially for the GPUs with lower computation capability. Meanwhile, as the number of MECs and VR users scaling up, the centralized DRL can become inefficient due to the following issues. First, the training time is bound by the gradient computation time, and the frequency of parameter updating grows linearly with the number of MECs and VR users. Second, as the frequency of parameter updating grows, it could potentially slow down the optimization process and result in problems with convergence \cite{Ddqn1}. 

Unlike the centralized DRL algorithm, the global objective in distributed DRL algorithm is the combination of each agent's local objective, and each agent needs to optimize its own objective. In the distributed DQN method, each agent learns independently from the other agents. When one of the agents selects an action based on the current state, the other agents can be approximated as part of the environment \cite{DQ_OFDMA}.

In our model, the central controller stores a copy of the model parameter $\pmb{\theta}_{\text{DDQN}}$. The $i$th MEC obtains the latest model parameter $\pmb{\theta}_{\text{DDQN}}$ from the central controller with $\widetilde{\pmb{\theta}}_{i} = \pmb{\theta}_{\text{DDQN}}$. Based on the observed state $S_{t}^{i}$, it will select an action $A_{t}^{i}$ in all available actions in $\mathcal{A}^i$. As a result, the environment will make a transition to the new state $S_{t+1}^{i}$ and a reward $R_{t}^{i}$ will be generated and fed back to the $i$th MEC. During training, the parameter $\widetilde{\pmb{\theta}}_{i}$ of the $i$th MEC can be updated as 
\begin{equation}
    \widetilde{\pmb{\theta}}_{i}^{t+1} = \widetilde{\pmb{\theta}}_{i}^{t} - \lambda_{\text{DDQN}}\nabla L_{i}(\widetilde{\pmb{\theta}}_{i}^{t}),
\end{equation}
where $\lambda_{\text{DDQN}}\in(0,1]$ is the learning rate, $L_{i}(\widetilde{\pmb{\theta}}_{i}^{t})$ is the loss function of the $i$th MEC, which can be denoted as
\begin{equation}
    L_i(\widetilde{\pmb{\theta}}_{i}^{t}) =(\hat{V}_{\text{DDQN}}-{V}_{\text{DDQN}}(S_{j}^{i},A_{j}^{i};\widetilde{\pmb{\theta}}_{i}^{t}))^{2},
\end{equation}
where
\begin{equation}
    \hat{V}_{\text{DDQN}} = R_{j+1}^{i} + \gamma\max_{A^{i}}{V}_{\text{DDQN}}(S_{j+1}^{i},A^{i};\bar{\widetilde{\pmb{\theta}}}_{i}^{t}).
\end{equation}
$(S_{j}^{i},A_{j}^{i},S_{j+1}^{i},r_{j+1}^{i})$ are randomly selected previous samples for $j\in\{t-M_r,..,t\}$ of the $i$th MEC. $\bar{\widetilde{\pmb{\theta}}}_{i}^{t}$ is the target Q-network which is used to estimate the future value of the state-action value function in the update rule. Furthermore, through deriving the loss function in (32) and computing the expectation of the selected samples, ${V}_{\text{DDQN}}^{*}(S^{i},A^{i})$ can be obtained. In addition, the updated parameter $\widetilde{\pmb{\theta}}_{i}$ of the $i$th MEC will be transmitted to the central controller and the model parameter $\pmb{\theta}_{\text{DDQN}}$ can be updated as
\begin{equation}
    \pmb{\theta}_{\text{DDQN}} = \frac{1}{K_{\text{DDQN}}^{\text{MEC}}}\sum\limits_{i=1}^{K_{\text{DDQN}}^{\text{MEC}}}\widetilde{\pmb{\theta}}_{i},
\end{equation}
where $K_{\text{DDQN}}^{\text{MEC}}$ is the number of the MECs associated with the VR user groups.

\begin{figure}[ht]
    \centering
    \includegraphics[width=3.5 in]{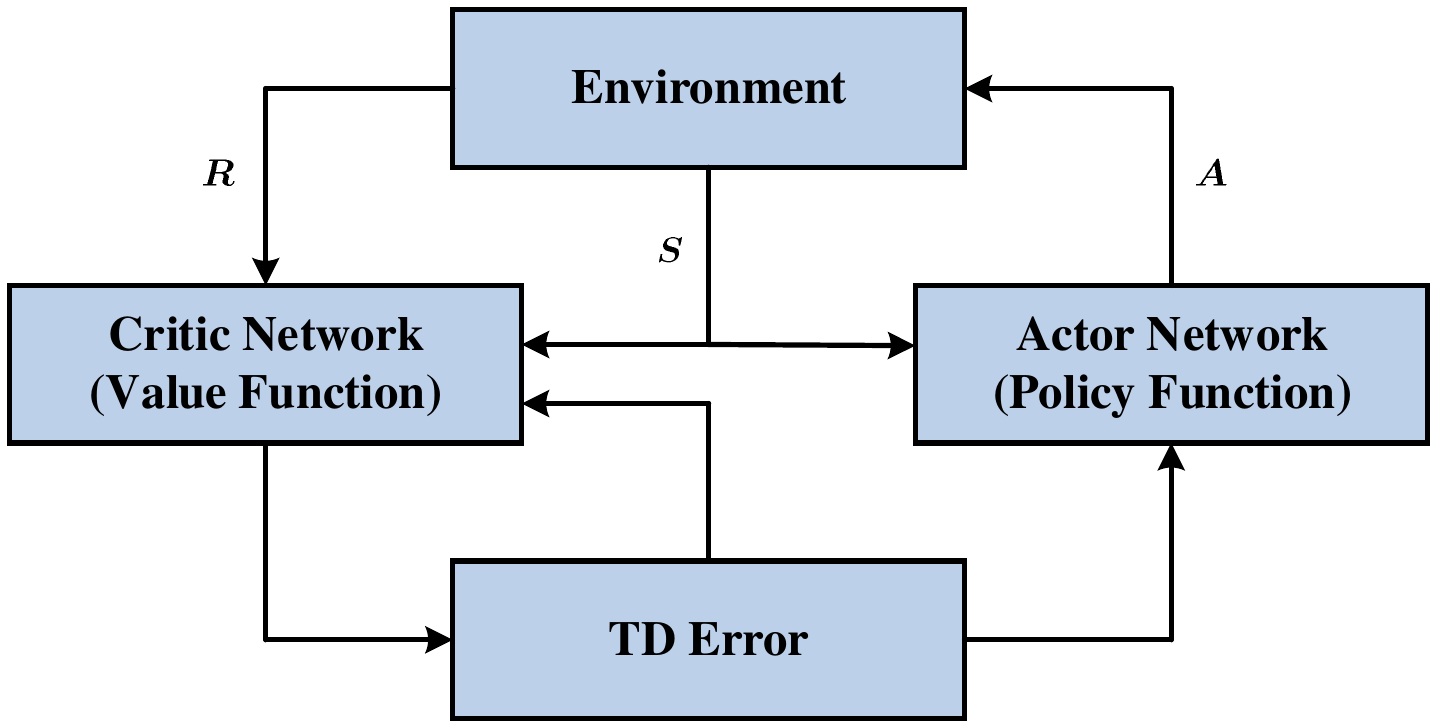}
    \caption{The Actor-Critic diagram of the MEC rendering scheme.}
    \label{basic_modules}
\end{figure}

\subsubsection{Centralized AC }
In the DQN algorithm, the optimal policy of the MEC rendering scheme is indirectly obtained through optimizing the state-action value function. However, unlike the DQN algorithm, AC algorithm is able to directly optimize the policy of the MEC rendering scheme. 

The core idea of the AC algorithm is to combine the advantages of Q-learning (value-based function) and the policy-gradient (policy-based function) algorithms. Consequently, the fast convergence of the value-based function and the directness of the policy-based function are all taken into consideration \cite{AC1, AC2}. As shown in Fig. 7, the AC network is consisted of two independent networks, namely, an actor network and a critic network. Through learning the relationship between the environment and the rewards, the critic network is able to get the potential rewards of the current state. Then, the critic network will guide the actor network to select proper actions and update the actor network in each epoch. Therefore, the AC algorithm is usually developed as a two-time-scale algorithm, including critic updating step and actor updating step, which leads to slow learning efficiency. A parameterized policy $\pi(A_t|S_t;\pmb{\theta}_{\text{AC}})$ is learned to select actions according to the current environment state. Then, the critic network will obtain the reward feedback from the environment and use the state-value function $V_{\text{AC}}(S_t;\pmb{w}_{\text{AC}})$ to evaluate the performed action. Meanwhile, a time-difference (TD) error is generated to reflect the performance of the performed action.

In particular, after performing action $A_t$ based on $S_t$ with policy $\pi$, the critic network uses TD error to evaluate the action under the current state, which can be expressed as
\begin{equation}
    \delta_t = R_t + \gamma{V}_{\text{AC}}(S_{t+1};\pmb{w}_{\text{AC}}^{t}) - V_{\text{AC}}(S_t;\pmb{w}_{\text{AC}}^{t}).
\end{equation}
Then, $\pmb{w}_{\text{AC}}^{t}$ can be updated as
\begin{equation}
    \pmb{w}_{\text{AC}}^{t+1} = \pmb{w}_{\text{AC}}^{t} + \lambda_{\text{critic}}\delta_{t}\nabla_{\pmb{w}_{\text{AC}}}V_{\text{AC}}(S_t;\pmb{w}_{\text{AC}}^{t}).
\end{equation}
where $\lambda_{\text{critic}}\in(0,1]$ is the learning rate of the critic network.

Meanwhile, in the actor network, the policy gradient method is usually adopted, which directly selects actions via parameterized policy. The parameter $\pmb{\theta}_{\text{AC}}^{t}$ can be updated as
\begin{equation}
    \pmb{\theta}_{\text{AC}}^{t+1} = \pmb{\theta}_{\text{AC}}^{t} + \lambda_{\text{actor}}\delta_{t}\nabla_{\pmb{\theta}_{\text{AC}}} \log\pi(A_t|S_t;\pmb{\theta}_{\text{AC}}^{t}),
\end{equation}
where $\lambda_{\text{actor}}\in(0,1]$ is the learning rate of the actor network.

Correspondingly, the parameters in the actor and critic network will be iteratively updated to maximize the objective function. The detailed AC algorithm of MEC rendering scheme is proposed in Algorithm 2.

\begin{algorithm}[t]
\begin{algorithmic}[1]
\caption{Actor-Critic to dynamic decision-making and optimization of wireless VR system}
\STATE Initialize learning rate $\lambda_{\text{critic}}\in(0,1]$, $\lambda_{\text{actor}}\in(0,1]$ and discount factor $\gamma\in[0,1)$.
\STATE Initialize parameters $\pmb{\theta}_{\text{AC}}$ and $\pmb{w}_{\text{AC}}$ for the actor and critic network, respectively.
\STATE Input the network state $S$ of the MEC rendering scheme.
\FOR{t = 1,...,T}
    \STATE According to $\pi(A|S_t;\theta)$, select the action $A\in\mathcal{A}$.
    \STATE The selected MECs render the required FoVs and multicast/unicast them to VR users due to the selected action $A_t$.
    \STATE MECs calculate the immediate reward $R_t$ and obtain the environment state $S_{t+1}$.
    \STATE Store transition $(S_t,A_t,R_t,S_{t+1})$.
    \STATE Calculate TD error $\delta_t$ according to (35).
    \STATE Update the parameters $\pmb{w}_{\text{AC}}$ of the critic network via (36).
    \STATE Update the parameters $\pmb{\theta}_{\text{AC}}$ of the actor network via (37).
\ENDFOR
\end{algorithmic}
\end{algorithm}


\subsubsection{Distributed AC}
Unlike the centralized AC algorithm, the agent in the distributed AC algorithm performs action and obtains reward based on its own observed state. For the critic network in each agent, it shares its estimate of the value function with others through the central controller. While for the actor network in each agent, it performs individually without the need to infer the policies of others \cite{Zhang_dis_AC}.

In our model, the $i$th MEC obtains the latest critic model parameter $\pmb{w}_{\text{DAC}}$ from the central controller, and let its own critic parameter $\Bar{\pmb{w}}_{t}^{i} = \pmb{w}_{\text{DAC}}$. At the $t$th time slot, according to the current environment state $S_{t}^{i}$ obtained by the $i$th MEC, a parameterized policy $\pi_{i}(S_{t}^{i};\Bar{\pmb{\theta}}_{t}^{i})$ is learned to select action $A_{t}^{i}$. Then, the critic network in the $i$th MEC will receive the reward feedback by the environment and evaluate the state-value function $V_{\text{DAC}}(A_{t}^{i}|S_{t}^{i}; \Bar{\pmb{w}}_{t}^{i})$. Similarly, the TD error $\delta_{t}^{i}$ of the $i$th MEC can be calculated to judge the performance of the performed action $A_{t}^{i}$, and the parameter $\Bar{\pmb{w}}_{t}^{i}$ of the critic network of the $i$th MEC can be updated as
\begin{equation}
    \Bar{\pmb{w}}_{t+1}^{i} = \Bar{\pmb{w}}_{t}^{i} + \Bar{\lambda}_{\text{critic}}\delta_{t}^{i}\nabla_{\Bar{\pmb{w}}^{i}}V_{\text{DAC}}(S_t^{i};\Bar{\pmb{w}}_{t}^{i}),
\end{equation}
where $\Bar{\lambda}_{\text{critic}}\in(0,1]$ is the learning rate of the critic network, and
\begin{equation}
    \delta_t^{i} = R_t^{i} + \gamma{V}_{\text{DAC}}(S_{t+1}^{i};\Bar{\pmb{w}}_{t}^{i}) - V_{\text{DAC}}(S_t^{i};\Bar{\pmb{w}}_{t}^{i}).
\end{equation}
Furthermore, for the parameter $\Bar{\pmb{\theta}}_{t}^{i}$ of the actor network of the $i$th MEC, it can be updated via
\begin{equation}
    \Bar{\pmb{\theta}}_{t+1}^{i} = \Bar{\pmb{\theta}}_{t}^{i} +\Bar{\lambda}_{\text{actor}}\delta_{t}^{i}\nabla_{\Bar{\pmb{\theta}}^{i}}\log\pi(A_t^{i}|S_t^{i};\Bar{\pmb{\theta}}_{t}^{i}),
\end{equation}
where $\Bar{\lambda}_{\text{actor}}\in(0,1]$ is the learning rate of the actor network. In addition, the updated parameter $\Bar{\pmb{w}}_{t}^{i}$ in the critic network of the $i$th MEC will be sent to the central controller and the critic model parameter $\pmb{w}_{\text{DAC}}$ can be updated as
\begin{equation}
    \pmb{w}_{\text{DAC}} = \frac{1}{K_{\text{DAC}}^{\text{MEC}}}\sum\limits_{i=1}^{K_{\text{DAC}}^{\text{MEC}}}\Bar{\pmb{w}}^{i},
\end{equation}
where $K_{\text{DAC}}^{\text{MEC}}$ is the number of the MECs associated with the VR user groups. Correspondingly, the parameters in the actor and critic network will be iteratively updated to maximize the objective function.

\begin{figure}[!t]
	\centering
	\subfloat[]{\includegraphics[width=3.0in]{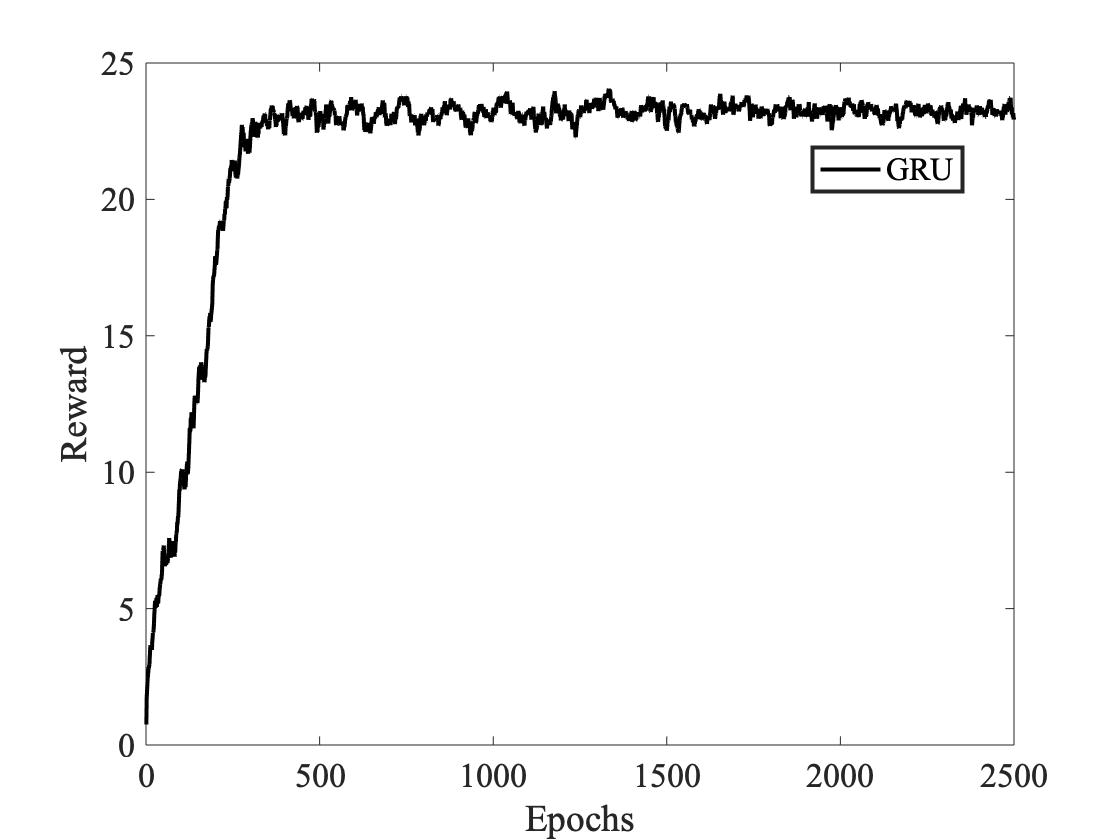}}\label{fig_first_case}
	\hfil
	\subfloat[]{\includegraphics[width=3.0in]{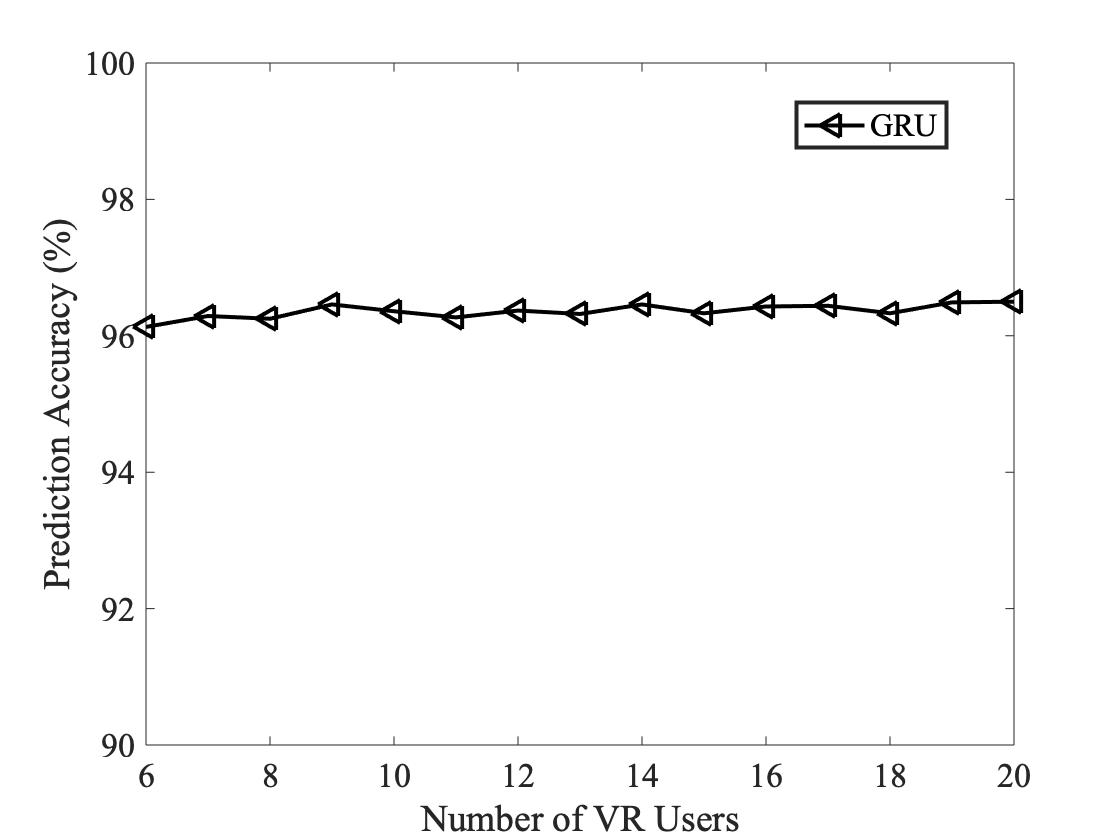}}\label{fig_second_case}
	\caption{(a) Total reward of FoV prediction of each epoch via GRU. (b) FoV prediction accuracy of GRU for varying number of VR users.}
	\label{basic_modules}
\end{figure}

\section{Simulation Results}
In this section, we examine the effectiveness of our proposed schemes with learning algorithms via simulation. For the learning algorithms, we set the learning algorithms use fully-connected neural network with two hidden layers and each layer has 128 ReLU units, we set the memory as 20, the minibatch size as 64, the learning rate for RNN as 0.005, the number of MECs as 8, the number of VR users as 8, $N_{\text{FoV}} = 8$, $D(t) = 3$, $\sigma^{2} = -110~\text{dBm}$, $\gamma = 0.9$, $\alpha = \beta = 3$, $\lambda_{\text{DQN}} = 0.05$, $\lambda_{\text{actor}} = 0.005$, $\lambda_{\text{critic}} = 0.05$, $T^{th} = 30 ~\text{ms}$, $\mathcal{R} = 1080\text{p}$, $\mathcal{C}_{\mathcal{R}} = 200$, $F_{max}^{\text{MEC}} = 5~\text{GHz}$, $F_{min}^{\text{MEC}} = 4~\text{GHz}$, $F^{\text{VR}} = 2~\text{GHz}$, $f^{\text{MEC}} = f^{\text{VR}} = 1000~ \text{Cycles/bit}$, and $R^{\text{fiber}} = 10~ \text{Gb/s}$. Consider a limited square area whose side length is 100 meters.

\subsection{FoV Prediction}
In the FoV prediction scheme, Brownian motion is deployed to simulate the eye movement of VR users. To obtain high accuracy in predicting FoV preference of each VR user in continuous time slots, a RNN model basd on GRU architecture is deployed. Fig. 8 (a) plots the total reward of FoV prediction of each epoch via RNN and Fig. 8 (b) plots the FoV prediction accuracy of RNN for varying number of VR users, respectively. It is observed that the prediction accuracy of the RNN remains $96\%$ despite the increasing number of VR devices. This is because RNN utilizes a memory window with length 20 to store with the input observations, which can capture the FoV preference of VR users in the past time slots.

\subsection{MEC Rendering Scheme}
Four DRL algorithms, including centralized DQN, distributed DQN, centralized AC, and distributed AC, are proposed to select proper MECs to render and transmit the required FoVs to VR users. For simplicity, we use ``w/ Pred", ``w/o Pred", ``w/ Migra", and ``w/o Migra" to represent ``with prediction", ``without prediction", ``with migration", and ``without migration" in the figures, respectively. To guarantee the fairness of each VR user, we use average QoE and VR interaction latency in the performance results.

\begin{figure}[ht]
    \centering
    \includegraphics[width=3.0 in]{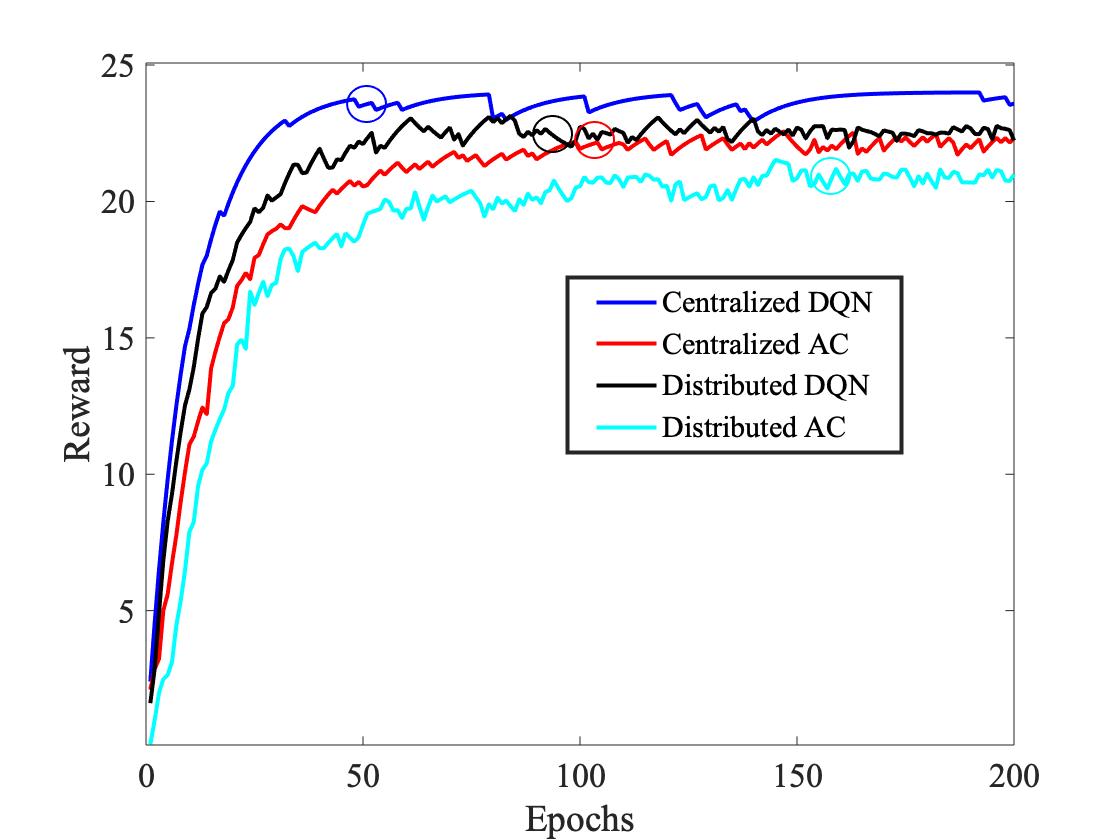}
    \caption{Total reward of the MEC rendering with prediction and migration scheme of each epoch via centralized/distributed DQN/AC learning algorithms.}
    \label{basic_modules}
\end{figure}

Fig. 9 plots the total reward of the MEC rendering with prediction and migration scheme of each epoch via centralized/distributed DQN/AC learning algorithms. Each result is averaged over 100 training trails. It is observed that the total reward and the convergence speed of these four DRL learning algorithms follows: Centralized DQN $>$ Distributed DQN $>$ Centralized AC $>$ Distributed AC. This is due to the experience replay mechanism and randomly sampling in DQN, which use the training samples efficiently and smooth the training distribution over the previous behaviours. As the model parameters in AC algorithm are updated in two steps, including critic step and actor step, the convergence speed of the AC algorithm is lower. Apparently, the convergence speed of the centralized learning algorithms is faster than that of the distributed learning algorithms. This is because the distributed learning needs more time to learn from each agent with only local observation and reward, whereas centralized learning can learn from global observations and rewards.

\begin{figure}[ht]
    \centering
    \includegraphics[width=3.0 in]{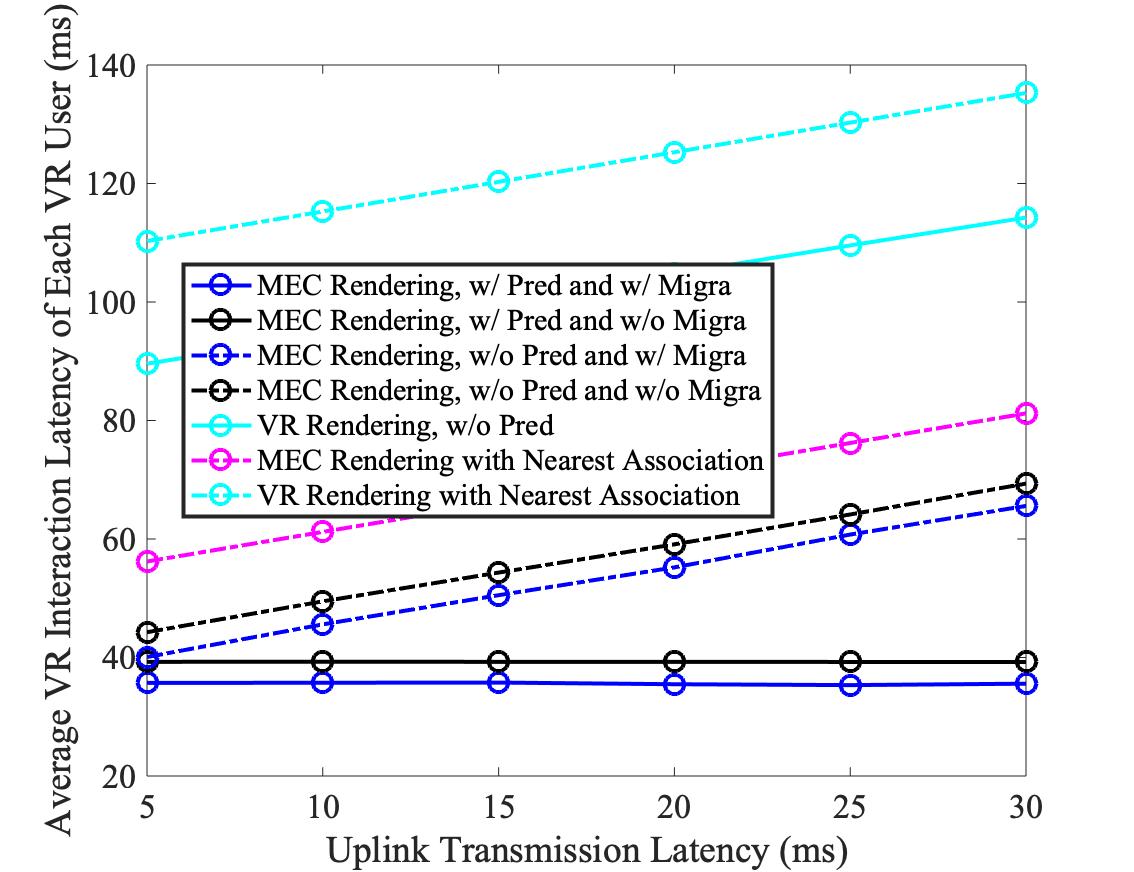}
    \caption{Average VR interaction latency of various MEC and VR rendering schemes via centralized DQN algorithm for varying uplink transmission latency.}
    \label{basic_modules}
\end{figure}

Fig. 10 plots the average VR interaction latency of various MEC and VR rendering schemes via centralized DQN algorithm for varying uplink transmission latency. We observe that all the MEC rendering schemes outperform that of the VR rendering schemes, with around $40~\text{ms}$ gain. This is because the processing ability of the MECs is much higher than that of the VR devices, and the data size of the FoV is smaller than that of the stitched 2D picture, which jointly decrease the rendering and downlink transmission latency. We also observe that the average VR interaction latency of the MEC rendering with prediction and migration scheme remains the same with increasing the uplink transmission latency, as the MECs do not need to wait for the uplink transmission of requested FoV from the VR devices before performing rendering.

In Fig. 10, we also compare our proposed learning-based schemes with those without learning. By comparing with the MEC/VR rendering scheme with nearest association scheme plotted using dash lines, we see our proposed learning-based MEC/VR rendering schemes achieve substantial gain in terms of VR interaction latency. This is due to that in the non-learning scheme, the VR user needs to transmit its requested FoV through uplink transmission and is always associated with the nearest MEC. Thus, it is possible that the MEC with low processing ability is selected to render the required FoV, which can increase the rendering latency.

\begin{figure}[ht]
    \centering
    \includegraphics[width=3.0 in]{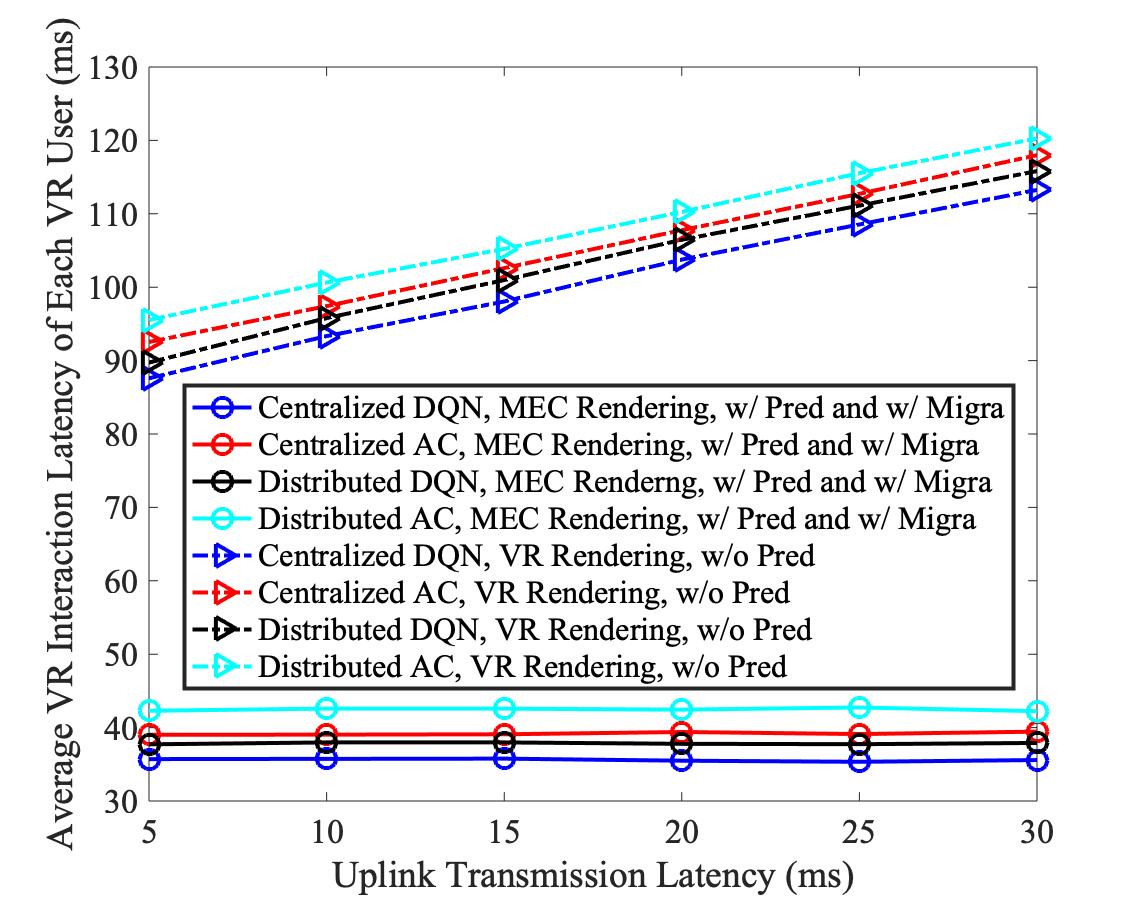}
    \caption{Average VR interaction latency of the MEC rendering with prediction and migration scheme and the VR rendering scheme via centralized/distributed DQN/AC learning algorithms for varying uplink transmission latency.}
    \label{basic_modules}
\end{figure}

Fig. 11 plots the average VR interaction latency of the MEC rendering with prediction and migration scheme and the VR rendering scheme via centralized/distributed DQN/AC learning algorithms for varying uplink transmission latency. It is observed that for the MEC rendering scheme achieves much lower latency (about $40~\text{ms}$) compared to VR rendering scheme. It is also seen that for the same rendering scheme, either the MEC or VR, the centralized DQN algorithm can achieve the minimum average VR interaction latency. This can be explained by the fact that the centralized learning algorithm learns a single policy common to the whole wireless VR system based on the global observations, while in the distributed learning algorithm, each agent only learns its own policy based on local observation.

\begin{figure}[ht]
	\centering
	\subfloat[]{\includegraphics[width=3.0in]{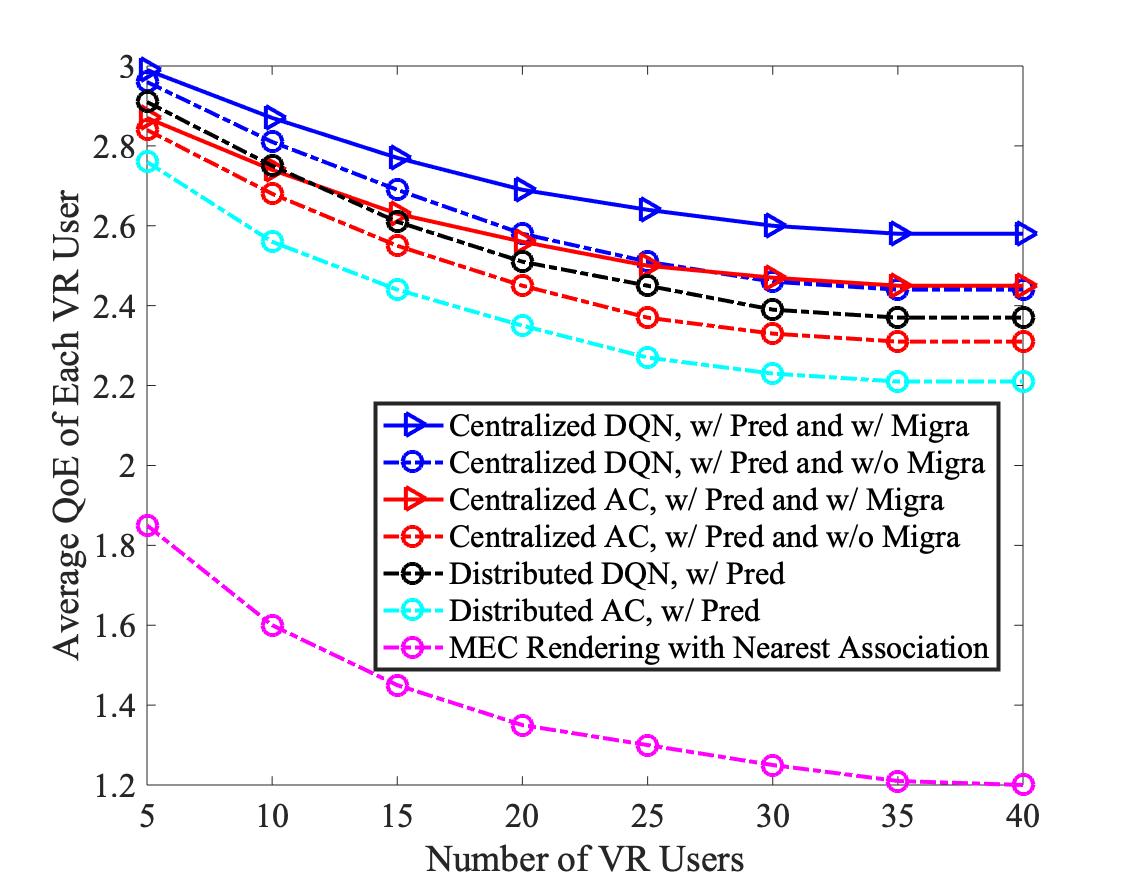}}\label{fig_first_case}
	\hfil
	\subfloat[]{\includegraphics[width=3.0in]{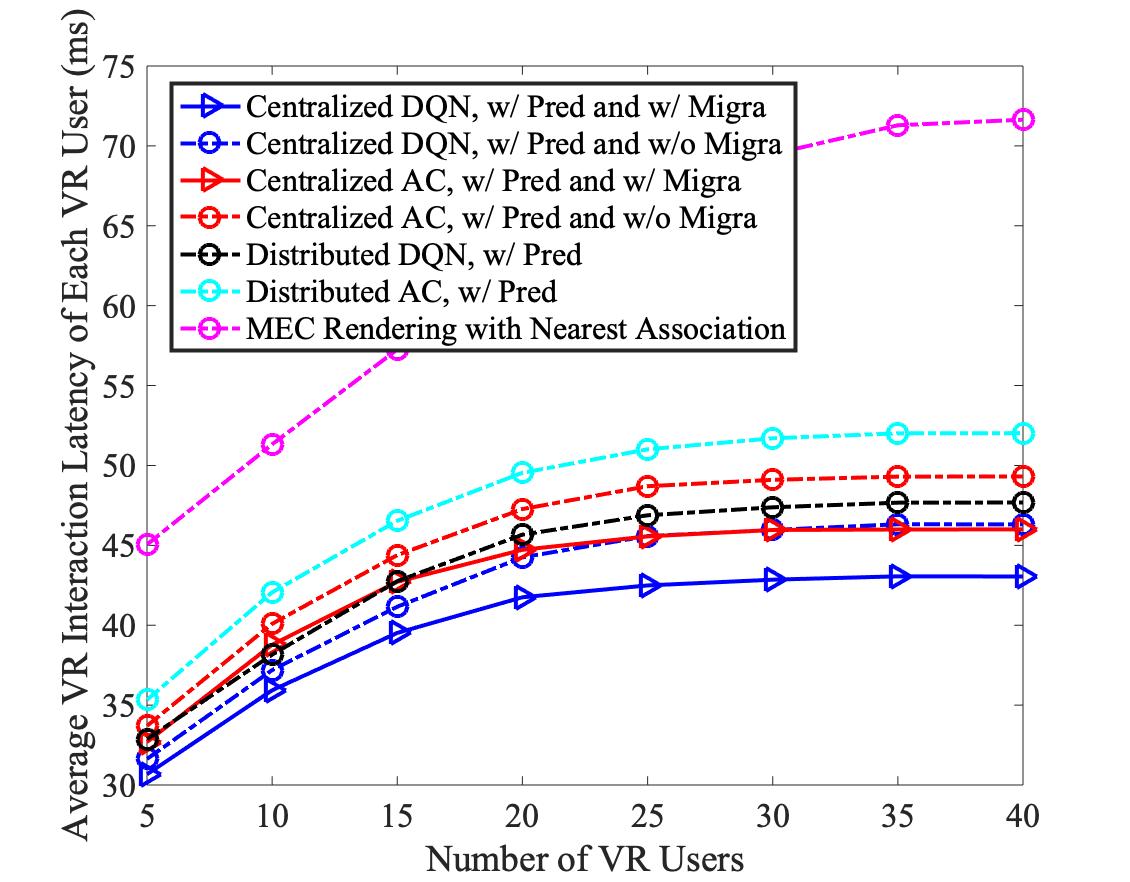}}\label{fig_second_case}
	\caption{Average QoE and VR interaction latency of MEC rendering with prediction with/without migration schemes via centralized/distributed DQN/AC learning algorithms with increasing number of VR users.}
	\label{basic_modules}
\end{figure}

Fig. 12 plots the average QoE and VR interaction latency of MEC rendering with prediction with/without migration schemes via centralized/distributed DQN/AC learning algorithms with increasing number of VR users, respectively. With increasing the number of VR users, the average QoE of VR user first decreases then becomes nearly stable as shown in Fig. 12 (a), whereas the average VR interaction latency first increases, then becomes nearly stable as shown in Fig. 12 (b).  This is because with increasing number of VR users, more MECs are activated to provide the downlink transmission for more requested different FoVs, which increase the interference among those transmission. As the number of VR users becomes too large, all the MECs become active to serve all VR users to render most of FoVs, the rendering latency and interference among VR users become stable.

Interestingly, we notice that for both centralized DQN and AC algorithms, we can see the performance gain of the MEC rendering with migration scheme over that without migration scheme in Fig. 12 (a) and (b). This is because the MECs with higher computing ability will be selected to render the same required FoV for migration, which decreases the rendering latency. Importantly, all the learning-based MEC rendering with prediction schemes substantially outperform the conventional non-learning based MEC rendering with nearest association scheme.

\begin{figure}[ht]
	\centering
	\subfloat[]{\includegraphics[width=3.0in]{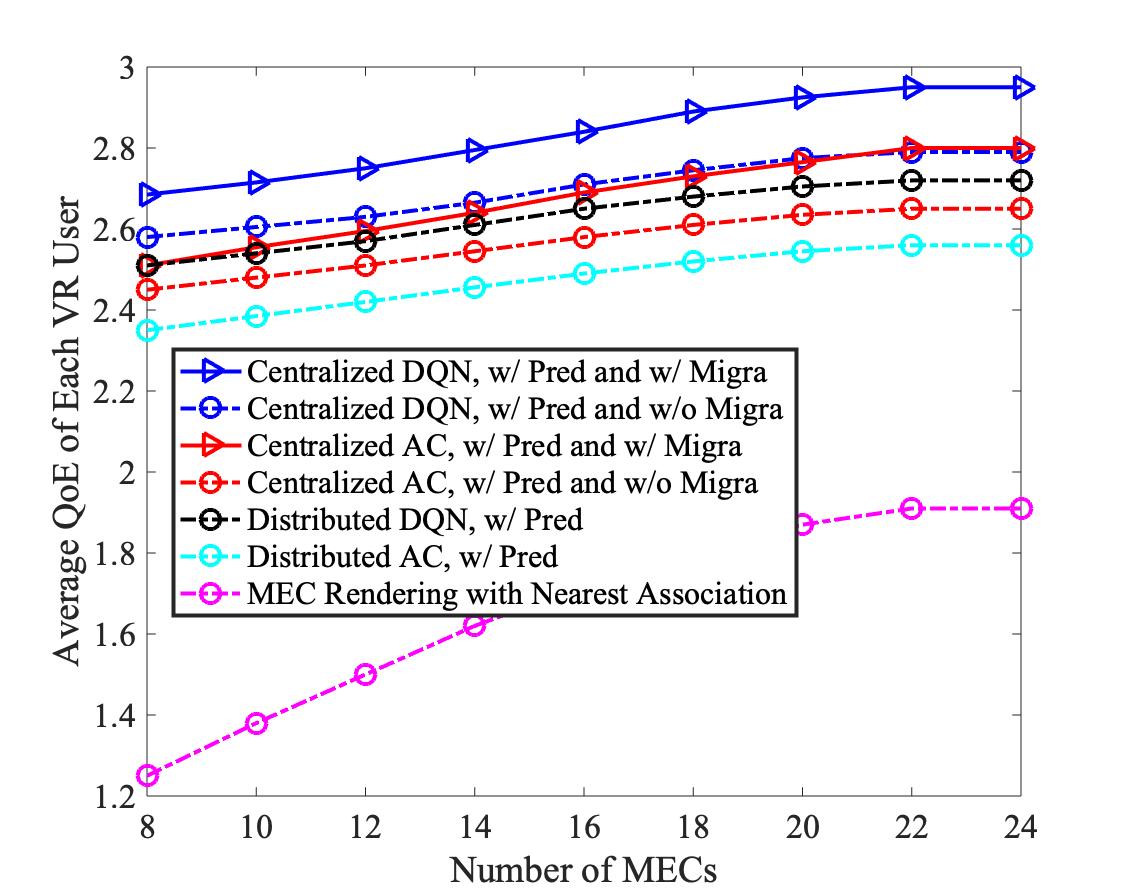}}\label{fig_first_case}
	\hfil
	\subfloat[]{\includegraphics[width=3.0in]{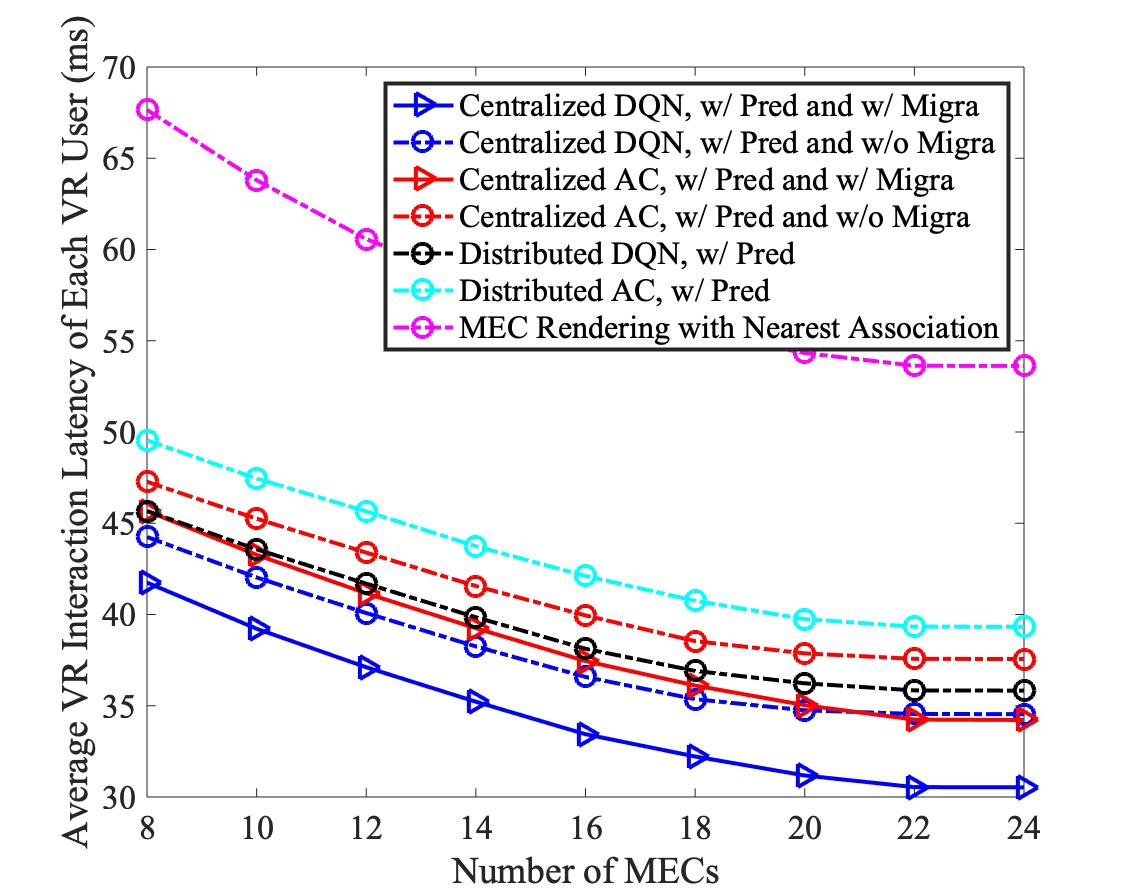}}\label{fig_second_case}
	\caption{Average QoE and VR interaction latency of MEC rendering with prediction with/without migration schemes via centralized/distributed DQN/AC learning algorithms with increasing number of MECs.}
	\label{basic_modules}
\end{figure}

Fig. 13 plots the average QoE and VR interaction latency of MEC rendering with prediction with/without migration schemes via centralized/distributed DQN/AC learning algorithms with increasing number of MECs, respectively. With increasing the number of MECs, the average QoE of VR user first increases then becomes nearly stable as shown in Fig. 13 (a), whereas the average VR interaction latency first decreases, then becomes nearly stable as shown in Fig. 13 (b). This is because as the number of MECs increases, the VR users will have more MEC choices to be selected, thus, nearer MECs with higher execution ability can be utilized to render the required FoVs, which reduces the rendering and downlink transmission latency. However, as the number of MECs becomes too large,  all MECs may be activated for rendering and downlink transmission, which leaves little gain for improvement.



\section{Conclusions}
In this paper, a decoupled learning strategy was developed to optimize real-time VR video streaming in wireless network, which considered FoV prediction and rendering MEC association. Specifically, based on GRU architecture, a RNN model was used to predict FoV preference of each VR user over time. Then, based on the correlation between the location and predicted FoV request of VR users, centralized and distributed DRL strategies were proposed to determine the optimal association between MEC and VR user group, and optimal rendering MEC for model migration, so as to maximize the long-term QoE of VR users. Simulation results shown that our proposed MEC rendering with prediction and migration scheme based on RNN and DRL algorithms substantially improved the long-term QoE of VR users and the VR interaction latency.

%





\ifCLASSOPTIONcaptionsoff
  \newpage
\fi





\bibliographystyle{IEEEtran}
\bibliography{IEEEabrv,Ref,ReferencesMP}
%
%

\end{document}